\documentclass[twocolumn,english,aps,prb,twocolumn,superscriptaddress,bibnotes,amsmath,amssymb,floatfix]{revtex4-1}
\usepackage[colorlinks=true,citecolor=blue,linkcolor=magenta]{hyperref}

\usepackage[flushleft]{threeparttable}
\usepackage[table]{xcolor}

\usepackage[utf8]{inputenc}
\usepackage[english]{babel}
\usepackage{amsmath,amsfonts,amssymb}
\usepackage[T1]{fontenc}
\usepackage{url}
\usepackage{soul}
\usepackage{changes}

\usepackage{amsmath}
\usepackage{amsfonts}
\usepackage{amssymb}
\usepackage[version=3]{mhchem}
\usepackage{stmaryrd}
\usepackage{epstopdf}
\usepackage{graphicx}
\graphicspath{{./Figures/}}

\definecolor{mygray}{gray}{0.95}

\begin{document}

\title{Optically reconfigurable quasi-phase-matching in silicon nitride microresonators}

\author{Edgars Nitiss}
\thanks{These authors contributed equally to the work.}
\affiliation{{\'E}cole Polytechnique F{\'e}d{\'e}rale de Lausanne, Photonic Systems Laboratory (PHOSL), STI-IEL, Station 11, Lausanne CH-1015, Switzerland.}

\author{Jianqi Hu}
\thanks{These authors contributed equally to the work.}
\affiliation{{\'E}cole Polytechnique F{\'e}d{\'e}rale de Lausanne, Photonic Systems Laboratory (PHOSL), STI-IEL, Station 11, Lausanne CH-1015, Switzerland.}

\author{Anton Stroganov}
\affiliation{LIGENTEC SA, EPFL Innovation Park, Bâtiment L, 1024 Ecublens, Switzerland.}

\author{Camille-Sophie Br\`es}
\email[]{camille.bres@epfl.ch}
\affiliation{{\'E}cole Polytechnique F{\'e}d{\'e}rale de Lausanne, Photonic Systems Laboratory (PHOSL), STI-IEL, Station 11, Lausanne CH-1015, Switzerland.}

\maketitle

\noindent\textbf{\noindent
Bringing efficient second-order nonlinear effects in integrated photonics is an important task motivated by the prospect of enabling all possible optical functionalities on chip. Such task has proved particularly challenging in silicon photonics, as materials best suited for photonic integration lack second-order susceptibility ($\chi^{(2)}$). Methods for inducing effective $\chi^{(2)}$ in such materials have recently opened new opportunities. 
Here, we present optically reconfigurable quasi-phase-matching in large radius Si$_3$N$_4$ microresonators resulting in mW level on-chip second-harmonic generated powers. Most importantly we show that such all-optical poling can occur unconstrained from intermodal phase-matching, leading to widely tunable second-harmonic generation. 
We confirm the phenomenon by two-photon imaging of the inscribed $\chi^{(2)}$ grating structures within the microresonators as well as by dynamic tracking of both the pump and second-harmonic mode resonances. 
These results unambiguously establish that the photogalvanic effect, responsible for all-optical poling, can overcome phase mismatch constraints even in resonant systems, and simultaneously allow for the combined record of output power and tunability. 
}

\section*{Introduction} 

The demonstration of efficient parametric frequency conversion on-chip has long been one of the main goals in integrated photonics\cite{leuthold2010nonlinear,eggleton2011chalcogenide,moss2013new,hausmann2014diamond,gaeta2019photonic,kues2019quantum}. In particular, embracing second-order nonlinearity ($\chi^{(2)}$) in integrated devices could provide means for higher conversion efficiency (CE) and increased functionality in applications related to metrology, quantum photonics and optical communications. 
To date, lithium niobate (LN)\cite{wang2018ultrahigh,wolf2018quasi,chen2019ultra, lu2019periodically,lu2020toward} and some III-V materials (AlN\cite{guo2016second,bruch201817}, GaN\cite{xiong2011integrated,roland2016phase}, GaAs\cite{kuo2014second,chang2019strong}, AlGaAs\cite{mariani2014second}, GaP\cite{lake2016efficient,logan2018400}, etc) are the most common waveguide material platforms that naturally possess $\chi^{(2)}$ nonlinearity. 
Yet, despite the tremendous development of these emerging platforms in the past years, they are still far from being optimal options given the fabrication complexity, device capability, and cost. 
Alternatively, silicon photonics has matured substantially towards low-cost and high-volume manufacturing, and holds the promise for CMOS-compatible co-integration with electronics. However, the typically employed materials, i.e. silicon, silicon nitride (Si$_3$N$_4$), and silica (SiO$_2$), lack $\chi^{(2)}$ response due to being either centrosymmetric or amorphous. To this end, considerable efforts have been made to induce effective $\chi^{(2)}$ in silicon photonic materials. These include symmetry-breaking in silicon waveguides mediated by externally applied electric field\cite{timurdogan2017electric} or charged defects\cite{castellan2019origin},
intrinsic symmetry-breaking at an SiO$_2$ interface\cite{zhang2019symmetry}, and recently all-optical poling (AOP) of Si$_3$N$_4$ waveguides\cite{billat2017large,porcel2017photo,hickstein2019self,nitiss2019formation,nitiss2020highly,nitiss2020broadband}. Si$_3$N$_4$ is of particular interest owing to its excellent linear and nonlinear properties\cite{blumenthal2018silicon}. 
Indeed, Si$_3$N$_4$-based optical devices have been widely used for linear signal processing\cite{roeloffzen2013silicon}, as well as various nonlinear applications like supercontinuum\cite{guo2018mid} and Kerr comb\cite{gaeta2019photonic} generation based on the third-order nonlinearity ($\chi^{(3)}$). 
The possibility to exploit $\chi^{(2)}$ processes adds essential new functionalities to the already strong portfolio of Si$_3$N$_4$\cite{hickstein2019self}.

%%%%%%%%%%%%%%%%%%%%%%%%%%%%%%%%%%%%%%%%%%%
The photo-induced $\chi^{(2)}$ nonlinearity in Si$_3$N$_4$ waveguides is generally initiated by a high power pulsed laser source\cite{billat2017large,porcel2017photo,hickstein2019self,nitiss2019formation,nitiss2020highly}. 
While this approach allows for broadly tunable second-harmonic (SH) generation, it remains less efficient compared to non-centrosymmetric counterparts mentioned above.
As such, a doubly resonant configuration, as explored in many other platforms\cite{wolf2018quasi,chen2019ultra, lu2019periodically,lu2020toward,guo2016second,bruch201817,xiong2011integrated,roland2016phase,kuo2014second,chang2019strong,mariani2014second,lake2016efficient,logan2018400}, has recently been adapted for SH generation in Si$_3$N$_4$\cite{levy2011harmonic,lu2020efficient}.   
% To further enhance the CE and output second-harmonic (SH) power as well as facilitate AOP, doubly resonant microresonator was recently explored in Si$_3$N$_4$\cite{lu2020efficient} as has been shown in many other platforms\cite{wolf2018quasi,chen2019ultra, lu2019periodically,lu2020toward,guo2016second,bruch201817,xiong2011integrated,roland2016phase,kuo2014second,chang2019strong,mariani2014second,lake2016efficient,logan2018400}. 
The use of a high quality factor (Q) Si$_3$N$_4$ microresonator not only facilitates the onset of optical poling, but also achieves ultra-high CE on-chip\cite{lu2020efficient} benefiting from the double enhancements both at pump and SH. 
In addition to the energy conservation, it is well known that phase-matching of the pump and SH waves also needs to be fulfilled. 
Generally, intermodal phase-matching (perfect phase-matching)\cite{guo2016second,bruch201817,xiong2011integrated,roland2016phase,lu2020efficient} or $\overline{4}$-quasi-phase-matching\cite{kuo2014second,chang2019strong,mariani2014second,lake2016efficient,logan2018400} are mostly employed in such doubly resonant configurations. For both cases, the azimuthal mode numbers of the participating pump and SH resonances have to be carefully matched, thus being subjected to very limited reconfigurability. 
An alternative approach is quasi-phase-matching (QPM) mediated by periodic domain inversion, which has been recently implemented in LN microresonators\cite{wolf2018quasi,chen2019ultra, lu2019periodically,lu2020toward}. Yet such option involves additional complex fabrication steps while the still remains a fixed and narrowband operation. In Si$_3$N$_4$ microresonators, AOP was demonstrated at a single resonance restricted by perfect phase-matching condition\cite{lu2020efficient}. This inevitably imposes stringent constraints at the design stage of the microresonators. 

% FIGURE 1 with AOP setup
\begin{figure*}[htp]
  \centering{
  \includegraphics[width = 1.0\linewidth]{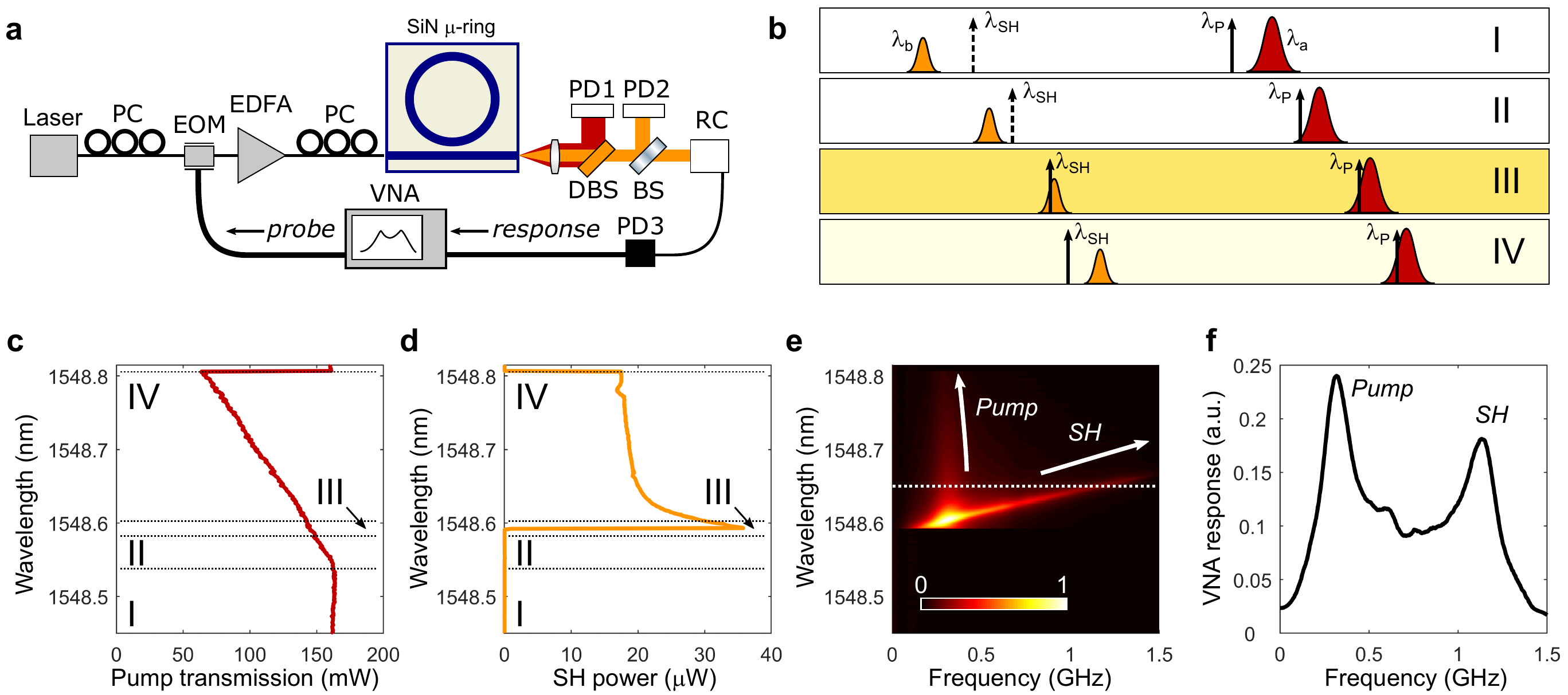}
  } 
    \caption{\noindent\textbf{All-optical poling of Si$_3$N$_4$ microresonators.} \textbf{a} Experimental setup for all-optical poling (AOP) of microresonators, and probing the detunings of pump and second-harmonic (SH) resonances. PC: polarization controller; EOM: electro-optic modulator;  EDFA: erbium-doped fiber amplifier; DBS: dichroic beamsplitter; BS: beamsplitter; RC: reflective collimator; PD1 and PD2: photodetectors for pump and SH light, respectively; PD3: $1~{\rm GHz}$ silicon photodetector; VNA: vector network analyzer. The sample temperature can be varied using a PID controller and a Peltier module. 
    \textbf{b} Schematic illustration of AOP process.
    $\lambda_{\rm P}$: pump wavelength; $\lambda_{\rm SH}$: SH wavelength; $\lambda_{\rm a}$: pump resonance wavelength; $\lambda_{\rm b}$: SH resonance wavelength. 
    ($\rm I$) Neither pump nor SH are resonant. 
    ($\rm II$)
    As $\lambda_{\rm P}$ is tuned into pump resonance $\lambda_{\rm a}$ (resonant for pump), both $\lambda_{\rm a}$ and $\lambda_{\rm b}$ red-shift due to the thermal and Kerr effects, compensating the mismatch between the virtual $\lambda_{\rm SH}$ and $\lambda_{\rm b}$ (still not resonant for SH).
    ($\rm III$) Once being doubly resonant for pump and SH, efficient AOP takes place. A nonlinear grating is inscribed and SH at $\lambda_{\rm SH}$ is generated. ($\rm IV$) The SH generation is sustained albeit $\lambda_{\rm b}$ walks off from $\lambda_{\rm SH}$. 
    \textbf{c} Measured on-chip pump transmission and \textbf{d} generated SH power during AOP process. Fundamental TE mode at pump and $2^{\rm nd}$ TE mode at SH (SH2) are responsible for this AOP process, confirmed by grating imaging. 
    \textbf{e} Measured VNA response map during AOP process. 
    Two distinct traces are observed corresponding to the detunings of pump and SH resonances, respectively.
    \textbf{f} Measured VNA response at pump wavelength of $1548.65~{\rm nm}$ (dashed line in \textbf{e}) with a clear double peak feature. 
    All the results shown in \textbf{c-f} are measured in a $146~{\rm GHz}$ microresonator with pump wavelength tuned from $1548.45~{\rm nm}$ to $1548.82~{\rm nm}$.
  }
 \label{Figure1n}
\end{figure*}

In this paper, we show for the first time, that automatic QPM can actually be induced in doubly resonant Si$_3$N$_4$ microresonators. 
Here AOP is simply initiated by a continuous-wave (CW) pump with moderate power.
An initial weak SH wave (e.g. arising from intrinsic nonlinearity at interface and defects\cite{levy2011harmonic,zhang2019symmetry,koskinen2017enhancement}), is resonantly enhanced together with the pump, and seeds the photogalvanic process. Similar to AOP in optical fibers\cite{osterberg1986dye, anderson1991model,dianov1995photoinduced, margulis1995imaging}, 
the interference of intracavity pump and SH gives rise to a self-organized space charge grating along the light propagation direction. This is confirmed experimentally by imaging the inscribed nonlinear grating structures with two-photon (TP) microscopy. The images also allow for the unambiguous identification of interacting SH modes based on the measured grating periods and shapes. We observe that multiple SH modes can participate in the AOP process without azimuthal mode numbers matching. Therefore, SH generation in AOP Si$_3$N$_4$ microresonator is unconstrained from the perfect phase-matching condition, significantly simplifying its design and allowing for unprecedented device performance based on QPM. With relatively small free spectral range (FSR) microresonators, we are able to optically reconfigure the $\chi^{(2)}$ gratings for SH generation in multiple pumped resonances. Our devices output mW level SH power with on-chip CE reaching several tens of $\%/{\rm W}$.
Moreover, we introduce an original technique that simultaneously probes pump and SH resonance detunings during the AOP event. 
We observe that, once the $\chi^{(2)}$ grating is inscribed, it is self-sustained and even enhanced when the pump is further detuned, albeit the SH resonance walks off. This enables remarkable SH wavelength tuning capability within one pump resonance.
The demonstrated results provide new insights towards bringing complete $\chi^{(2)}$ and $\chi^{(3)}$ nonlinearities on CMOS-compatible platforms.

\section*{Results} 
\noindent \textbf{AOP of Si$_3$N$_4$  microresonators.} Figure \ref{Figure1n}a shows the experimental setup for AOP of Si$_3$N$_4$ microresonators (see Methods). Light from a tunable CW laser is amplified, and coupled to Si$_3$N$_4$ microresonators using a lensed fiber. 
%The input coupling loss is estimated to be $2.4~{\rm dB}$. 
While AOP in Si$_3$N$_4$ microresonators can be realized in both TE and TM polarizations, %the AOP principle is similar in either condition. 
we chose to operate at TE excitation, as to facilitate TP imaging of the $\chi^{(2)}$ gratings since the charge separation takes place in the sample plane.
% because in such case the charge separation during grating inscription takes place in the sample plane making it easier to detect and characterize using TP imaging. 
The output pump and SH light from the chip are collected using a microscope objective, then separated and measured at their respective photodetectors (PD1 and PD2). 
In the experiment, we use two Si$_3$N$_4$ microresonators with radii of $158~{\mu \rm m }$ (FSR $\sim 146~{\rm GHz}$) and $119~{\mu \rm m}$ (FSR $\sim 196~{\rm GHz}$), and both of them exhibit loaded Qs $\sim 0.75\times 10^{6}$ for pump resonances around $1550~{\rm nm}$ (see Supplementary Note 1).
Since the microresonators were intentionally designed for light coupling at telecom-band using a single bus waveguide, the direct characterization of the relevant SH mode resonances is difficult. 
Nevertheless, this does not affect the occurrence of photo-induced QPM, while an additional bus waveguide may be introduced for efficient SH in/out coupling\cite{lu2020efficient}.  
% In future designs this limitation can be easily circumvented with the addition of a SH drop waveguide\cite{lu2020efficient}.
Instead, we propose an effective method for dynamic tracking of both the pump and SH resonance detunings while poling (see Methods). 
In this scenario, the laser is weakly phase-modulated in an electro-optic modulator (EOM) by a scanning microwave signal from a vector network analyzer (VNA). 
Then the generated signal at SH band is measured by a fast silicon photodector (PD3) to obtain the VNA response. 
The transfer function measured by the VNA is explicitly described in Supplementary Note 2, and is found to provide the detuning information of both interacting resonances.

Doubly resonant condition sets the prerequisite to initiate AOP, as illustrated in Figure \ref{Figure1n}b. 
The pump wavelength $\lambda_{\rm P}$ is initially set on the blue side of the cold pump resonance $\lambda_{\rm a}$, while the virtual SH wavelength $\lambda_{\rm SH} = \lambda_{\rm P}/2$ is expected to be on the red side of a particular SH mode resonance $\lambda_{\rm b}$ (region I). 
As the pump is red-detuned closer to the pump resonance, both pump and SH resonances would also be red-shifted due to the thermal and Kerr effects in the microresonator (region II). 
It is important to note that the SH resonance red-shifts faster than that of the pump\cite{lu2020efficient} (see Supplementary Note 1). 
As a consequence, the pump and SH resonances eventually match with $\lambda_{\rm P}$ and $\lambda_{\rm SH}$, respectively, i.e. the pump and SH become doubly resonant (region III).
A seed SH signal then efficiently initiates the photogalvanic effect, leading to photo-induced QPM and SH generation. 
A remarkable feature of the device manifests itself once the pump wavelength is further red-detuned.
Despite the walk-off of the SH resonance and deviation from the ideal QPM condition, efficient SH generation is still maintained (region IV).
The generated SH  and the intracavity $\chi^{(2)}$ grating form a dynamic equilibrium by means of self-sustaining feedback. 

A typical AOP process is experimentally demonstrated as shown in Figure. \ref{Figure1n}c-f. 
When the laser scans over a pump resonance from blue to red, we record the pump transmission and SH power on-chip (Figure \ref{Figure1n}c and d), as well as their resonance detunings (Figure \ref{Figure1n}e and f). 
Before the pump enters the pump resonance (region I) and before the seed SH light is resonantly enhanced (region II), no SH signal is generated. Once being doubly resonant (region III), we observe a sharp rise in SH signal, which indicates the AOP is initiated, here for pump wavelength around $1548.60~{\rm nm}$. 
For this particular resonance, the $\chi^{(2)}$ grating results from the interference of the fundamental TE mode at pump and the $2^{\rm nd}$ TE mode at SH (SH2) as inferred by TP imaging. 
The SH generation is then maintained until the pump is tuned out of its resonance (region IV). 
To gain insight into this behavior, we apply the aforementioned method to track the detunings of both resonances. 
Figure \ref{Figure1n}e illustrates the normalized VNA responses at different wavelengths. 
After AOP occurs, we observe two peaks in the measured transfer functions, which respectively correspond to pump and SH resonances (see Supplementary Note 2). In Fig. \ref{Figure1n}f we clearly visualize the double peak feature of the measured responses when plotted for a given pump wavelength (dashed line in Figure \ref{Figure1n}e). The peak closer to the DC response indicates the pump resonance, while the other peak indicates the SH resonance. Noticeably, the SH continues to be generated even when the SH resonance has greatly moved away from the SH wavelength. The VNA response maps for some other resonances are included in Supplementary Note 3. 

% FIGURE 2 Sweeps, grating illustration and imaging
\begin{figure*}[htbp]
  \centering{
  \includegraphics[width = 1\linewidth]{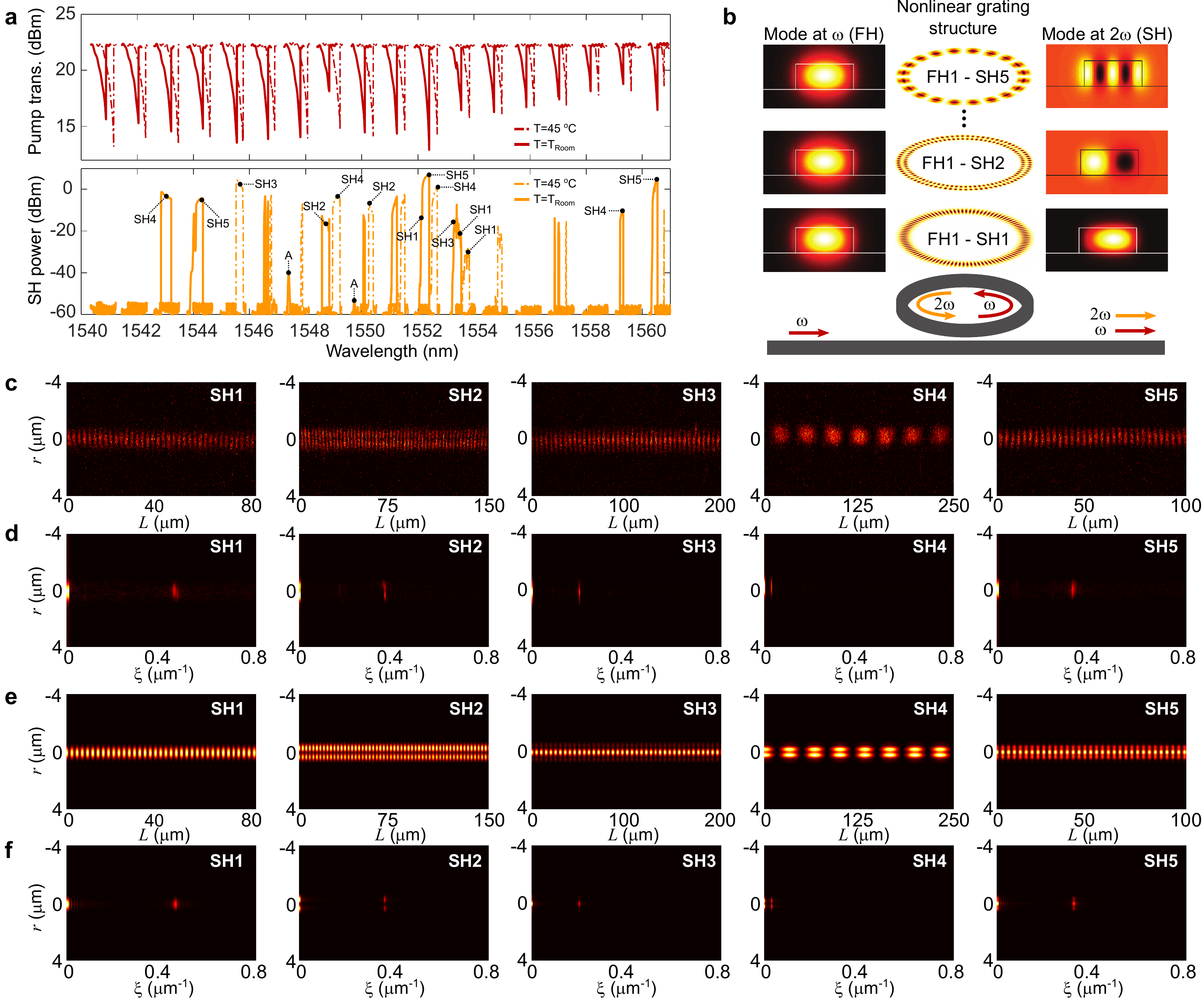}
  } 
  \caption{\noindent\textbf{Second-harmonic generation via reconfigurable quasi-phase-matched $\chi^{(2)}$ gratings.} 
  \textbf{a} On-chip pump transmission and generated SH power during the wavelength sweep in the $146~{\rm GHz}$ microresonator: without temperature control ($T=T{\rm_{Room}}$, solid) and with temperature stabilization ($T=45~{\rm ^\circ C}$, dashed). The interacting SH modes confirmed by two-photon (TP) imaging are indicated. 
  \textbf{b} Artistic representation of $\chi^{(2)}$ grating structures in the microresonator. 
  The periods and shapes of inscribed $\chi^{(2)}$ gratings are given by the interference between the fundamental TE mode at pump (FH1) and several TE modes at SH (SH1, SH2, and SH5 are depicted).
  \textbf{c} Experimentally retrieved $\chi^{(2)}$ gratings along the circumference of the microresonator.
%   Here arranged from left to right cases when fundamental pump mode with SH modes from fundamental to fifth are matched, respectively. 
  The grating structures (from left to right) are obtained at AOP wavelengths of $1552.15~{\rm nm}$ (SH1), $1548.80~{\rm nm}$ (SH2), $1553.38~{\rm nm}$ (SH3), $1559.33~{\rm nm}$ (SH4) and $1560.50~{\rm nm}$ (SH5).
  \textbf{d} Spatially-resolved Fourier analysis of  experimentally retrieved $\chi^{(2)}$ grating structures.
  \textbf{e} Simulated $\chi^{(2)}$ gratings along the circumference of the microresonator corresponding to the participating SH modes in \textbf{c}.
  \textbf{f} Spatially-resolved Fourier analysis of  simulated $\chi^{(2)}$ grating structures.
  }
 \label{Figure2n}
\end{figure*} 

\vspace{0.1cm}

\begin{figure*}[hbt!]
  \centering{
  \includegraphics[width = 0.95\linewidth]{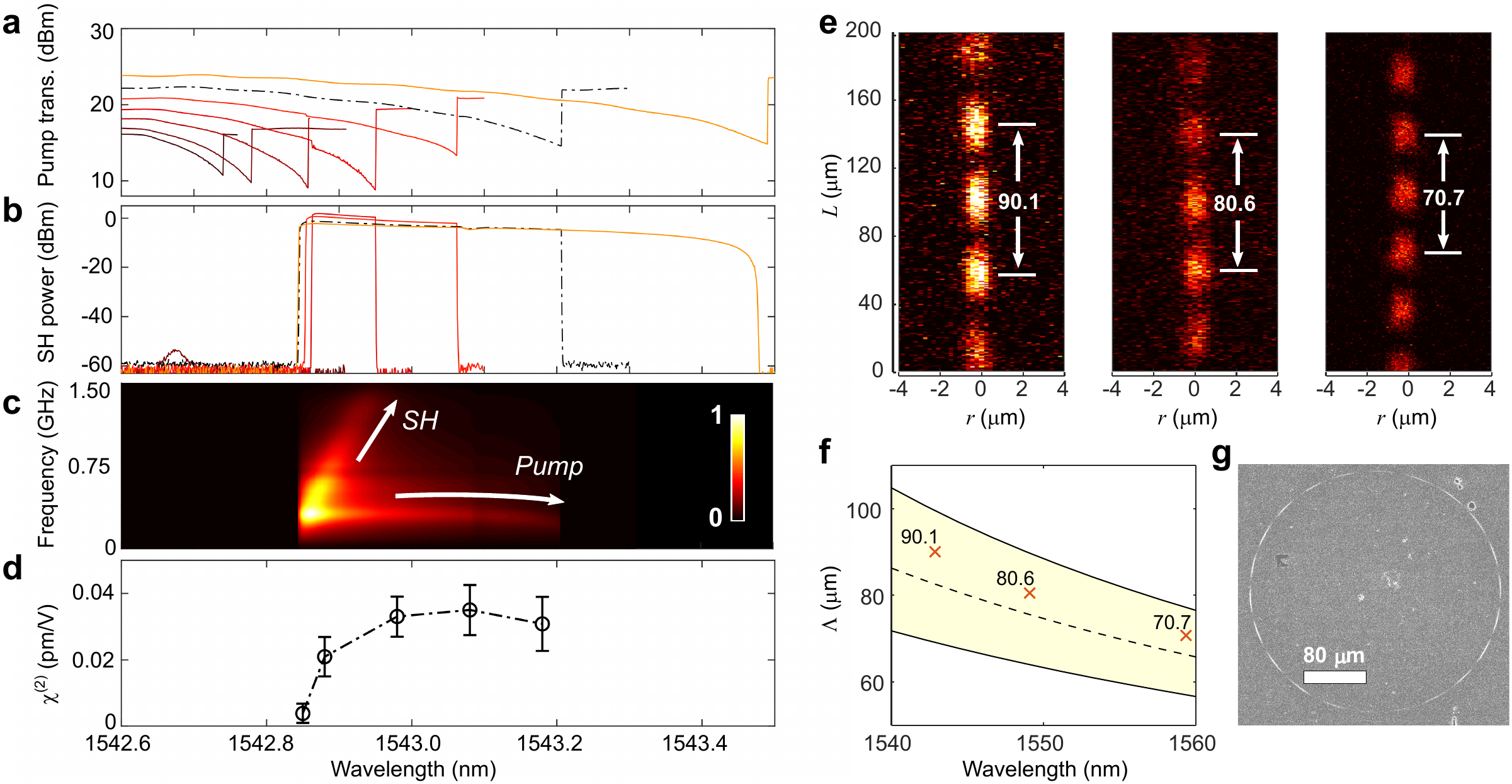}
  } 
  \caption{\noindent\textbf{Second-harmonic generation and $\chi^{(2)}$ grating characteristics based on the pump and SH4 mode interaction in the 146 GHz Si$_3$N$_4$ microresonator}. 
  \textbf{a} Pump transmission and \textbf{b}  Generated SH power on-chip during the wavelength sweeps over the pump resonance near $1542.8~{\rm nm}$, at different pump power levels.
  \textbf{c} VNA response map recorded with $22.2~{\rm dBm}$ pump power (correspond to the dashed lines in \textbf{a} and \textbf{b}).
  \textbf{d} $\chi{^{(2)}}$ characterization at different wavelengths with $22.2~{\rm dBm}$ pump power.
%   At highest coupled power the $10~{\rm dB}$ SH generation bandwidth is around $605~{\rm nm}$. 
  \textbf{e} Measured $\chi^{(2)}$ grating structures with periods of $90.1~{\rm \mu m}$, $80.6~{\rm\mu m}$ and $70.7~{\rm\mu m}$, obtained at $1542.90~{\rm nm}$, $1549.10~{\rm nm}$ and $1559.35~{\rm nm}$, respectively. 
  The corresponding numbers of grating periods inside the microresonator are $11$, $12$ and $14$, respectively. 
  \textbf{f} Simulation of the $\chi^{(2)}$ grating periods. Red crosses: experimentally retrieved $\chi^{(2)}$ grating periods in \textbf{e}; Dashed line: simulated $\chi^{(2)}$ grating period;
  Yellow region: simulated grating period variation considering device fabrication tolerances. 
  \textbf{g} Reconstructed TP image of the entire microresonator poled at $1543.0~{\rm nm}$, based on superposition of several depth-scanned TP images. 
  $11$ QPM grating periods are clearly recognized.}
 \label{Figure3n}
\end{figure*}

\noindent\textbf{Reconfigurable QPM and grating imaging.} 
Figure \ref{Figure2n}a illustrates versatile AOP of the $146~{\rm GHz}$ microresonator in the wavelength range between $1540~{\rm nm}$ to $1561~{\rm nm}$. 
When the wavelength sweep is performed without external temperature control ($T=T{\rm_{Room}}$), 11 out of the 18 available resonances show occurrences of AOP. By leveraging thermal control, the relative frequency offsets between the pump resonances and various SH mode resonances can be effectively modified. 
For example, when the chip holder temperature is stabilized at $T=45~{\rm ^oC}$, some additional resonances can support SH generation. Versatile AOP is enabled by the highly flexible doubly resonant condition between the fundamental pump mode and several SH modes. 
Such doubly resonant condition with detuning terms is given by:
\begin{equation}
\frac{2\pi n_{\rm eff, a (b)}}{\lambda_{\rm P(SH)}}2\pi R = 2\pi m_{\rm a (b)} - \theta_{\rm a (b)}
\label{doubly_resonant}
\end{equation}
where $R$ is the radius of the microresonator, $n_{\rm eff, a (b)}$, $m_{\rm a (b)}$, and $\theta_{\rm a (b)}$ are the effective refractive indices, azimuthal mode numbers, and phase offsets of pump and SH modes, respectively. For pump and SH wavelengths in the vicinity of their corresponding resonance, we have $|\theta_{\rm a (b)}/2\pi| \ll 1$.
The phase offsets can be expressed as $\theta_{\rm a (b)} =\delta_{\rm a (b)}/ \Delta f_{\rm FSR, a (b)}$, 
with $\delta_{\rm a (b)}$ the effective detunings in angular frequency and $\Delta f_{\rm FSR, a (b)}$ the FSRs at pump (SH) wavelengths. Besides being doubly resonant, phase-matching  condition is also required in the SH generation process. The phase-mismatch between the pump and SH writes:
\begin{equation}
\Delta k  
= \frac{2\pi n_{\rm{b,eff}}}{\lambda_{\rm SH}}- 
 \frac{ 4\pi n_{\rm{a,eff}}}{\lambda_{\rm{P}}} 
= \frac{m_{\rm b}- 2m_{\rm a}}{R} - \frac{\theta_{\rm b}- 2\theta_{\rm a}}{2\pi R}
\label{delta_k}
\end{equation}
where the second equality in Eq. \eqref{delta_k} is obtained using Eq. \eqref{doubly_resonant}. 
In AOP, such phase-mismatch is automatically compensated by the self-organized $\chi^{(2)}$ grating, i.e. $\Delta k = 2\pi/\Lambda$, with $\Lambda$ the grating period. 
The $\chi^{(2)}$ grating structure follows the interference of pump and SH fields\cite{baranova1990theory}:
\begin{equation}
\chi^{(2)}(\phi) \sim (E_{\rm P}^2)^\ast E_{\rm SH} \exp(i\Delta k R \phi) + {\rm c.c.}
\label{chi_2}
\end{equation}
where $E_{\rm P}$ and $E_{\rm SH}$ are the optical fields at pump and SH, respectively. $\phi$ denotes the azimuthal angle with reference to the center of a microresonator, and ${\rm c.c.}$ stands for complex conjugate. Notably, when phase offsets satisfy $\theta_{\rm b} = 2\theta_{\rm a}$, an integer number of grating periods $N=2\pi R/\Lambda = |m_{\rm b}- 2m_{\rm a}|$ is inscribed on the circumference of the microresonator.
Figure \ref{Figure2n}b showcases the artistic representation of 
inscribed QPM grating structures (not to the exact number of periods due to illustration purpose). 
They are based on the interactions between fundamental pump mode (FH1) and several SH modes, here shown for fundamental (SH1), $2^{\rm nd}$ (SH2), and $5^{\rm th}$ (SH5) with simulated spatial mode profiles (see Supplementary Note 1).
Regardless of the resonating SH mode in AOP, the QPM grating period $\Lambda$ automatically compensates the phase mismatch of pump and SH.
The AOP is thus initiated by the QPM grating without the need of exquisite perfect phase-matching ($m_{\rm b} = 2m_{\rm a}$)\cite{lu2020efficient}.

To reveal the $\chi^{(2)}$ gratings, we image the poled microresonators using a TP microscope (see Methods). 
Such technique has been previously applied to capture the $\chi^{(2)}$ response in poled optical waveguides\cite{hickstein2019self,nitiss2019formation}. 
We observe the formation of versatile $\chi^{(2)}$ gratings by performing TP imaging of microresonators poled at various resonances.
Figure \ref{Figure2n}c shows several retrieved grating structures along the circumference of the $146~{\rm GHz}$  microresonator, after coordinate transformation (see Supplementary Note 4). 
By spatially-resolved Fourier analysis, we acquire the spatial frequency graphs (Figure \ref{Figure2n}d) allowing for precise identification of the grating periods and shapes. 
The observed $\chi^{(2)}$ grating period is the result of the interaction between the fundamental pump mode and various SH modes. 
To unambiguously account for the participating SH modes, the grating periods and shapes as well as their spatial frequency graphs are simulated (see Supplementary Note 4) as displayed in Figure \ref{Figure2n}e and f.
By comparing the experimental and simulated grating images, we are able to identify the possible interacting SH modes from the fundamental SH mode (SH1) up to $5^{\rm th}$ SH mode (SH5). AOP via SH modes higher than SH5 is not experimentally observed, which may be attributed to their increased total losses, so that insufficient to initiate the photogalvanic effect.

Occasionally, we also observe two distinct $\chi^{(2)}$ gratings subsequently inscribed within the same pump resonances, e.g. the resonances near $1552.0~{\rm nm}$ and $1553.2~{\rm nm}$ as indicated in Figure \ref{Figure2n}a. For these cases, SH generation is initiated by resonant condition of one SH mode, and is then taken over by another SH mode with further laser detuning. This is verified by imaging the $\chi^{(2)}$ grating structures, and can also be inferred from the VNA measurement (see Supplementary Note 3).

\vspace{0.1cm}

\noindent\textbf{SH generation bandwidth and $\chi^{(2)}$ grating characteristics.} 
A remarkable feature of photo-induced SH generation in microresonators is its unusual bandwidth.
Figure \ref{Figure3n}a-d shows AOP of the $146~{\rm GHz}$ microresonator at a particular pump resonance near $1542.8~{\rm nm}$. The on-chip pump transmission (Figure \ref{Figure3n}a) and generated SH power (Figure \ref{Figure3n}b) are recorded at different pump power levels during the AOP event. 
As expected, the pump thermal triangle is prolonged at high pump power, due to the thermal-induced pump resonance drag. 
Moreover, once AOP is initiated, we observe the exceptionally broad spectral bandwidth over which SH generation can be maintained. 
At the highest pump power, we measure a $10~{\rm dB}$ SH generation bandwidth as large as $605 ~{\rm pm}$.
Figure \ref{Figure3n}c depicts the VNA map measured with a $22.2~{\rm dBm}$ on-chip pump power, which corresponds to the dashed curves in Figure \ref{Figure3n}a and b. 
% For AOP with a $22.2~{\rm dBm}$ on-chip pump power, we also perform the VNA measurement to probe the pump and SH detunings. 
Both pump and SH resonance detunings are clearly observed in the measured VNA responses. 
It confirms again that AOP can be effectively sustained despite a large walk-off of the SH resonance. 
To understand this behavior, we characterize the $\chi^{(2)}$ strength inside the microresonator at different detunings (see Supplementary Note 5), as shown in Figure \ref{Figure3n}d.
% The $\chi^{(2)}$ values at different pump wavelengths are plotted in Figure \ref{Figure3n}d. 
Counter-intuitively, we find that the measured $\chi^{(2)}$ is weakest right after the initiation of AOP, i.e. when pump and SH are doubly resonant. 
% when the SH and pump mode cavity resonances are matched, nevertheless, this condition still results in the highest measured SH power mainly due to the double resonant enhancement. 
When the SH resonance gradually walks off from the generated SH wavelength, the strength of $\chi^{(2)}$ is enhanced.
% At this point, the relationship between interacting mode cavity resonance detuning and resulting $\chi^{(2)}$ inside the resonator is still unclear. 
Then the inscribed $\chi^{(2)}$ reaches an equilibrium that depends on the magnitude of the photogalvanic current and waveguide material conductivity\cite{anderson1991model}. This explains in part the ultrabroad SH generation bandwidth in AOP process, yet the full theoretical model describing the doubly resonant AOP still needs further investigation.

% FIGURE 4 Poling resonance ratio, CEs
\begin{figure*}[hbt!]
  \centering{
  \includegraphics[width = 0.7\linewidth]{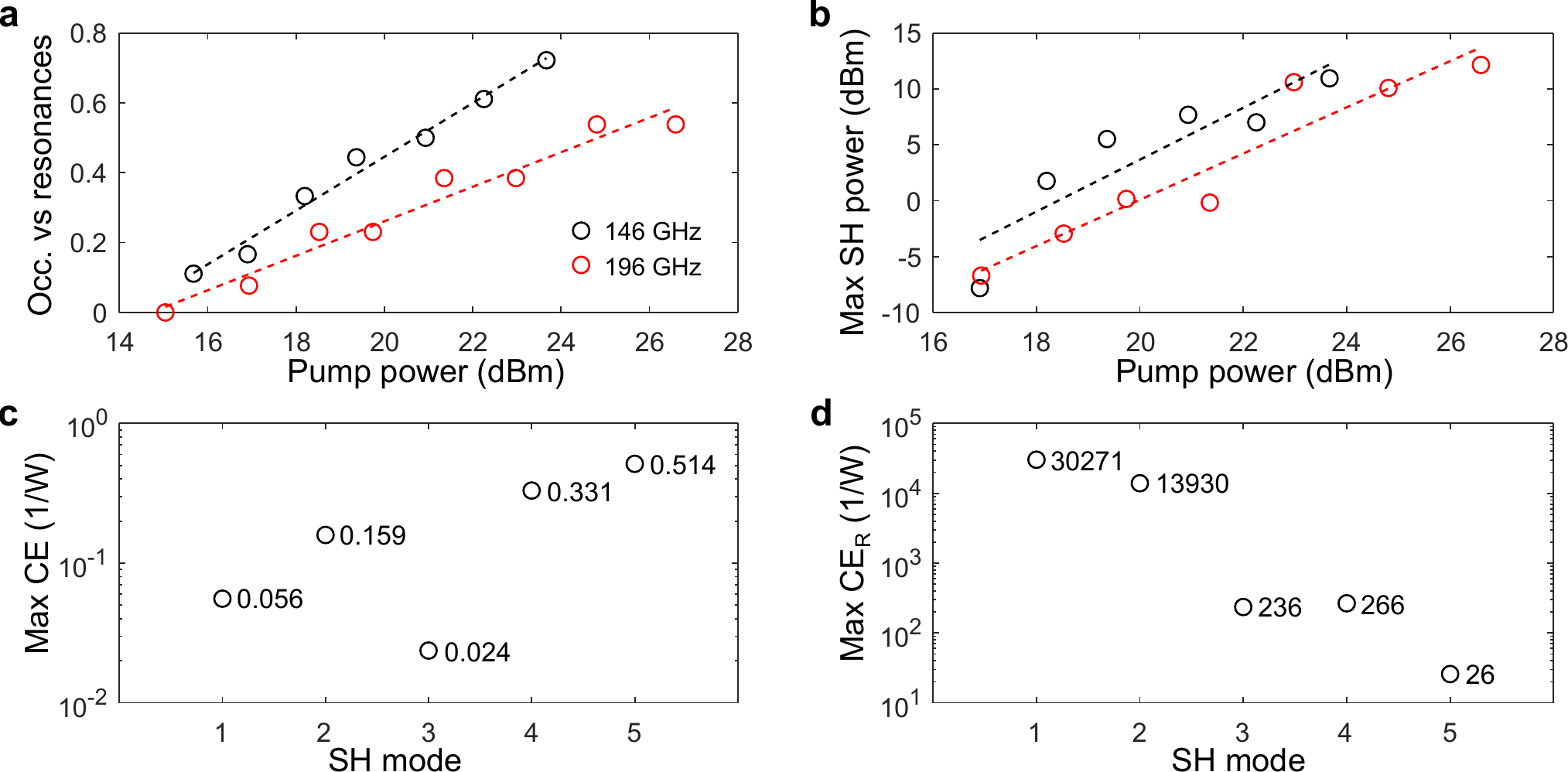}
  } 
  \caption{\noindent\textbf{Performance of second-harmonic generation in Si$_3$N$_4$ microresonators.} 
  \textbf{a} SH generation probability (AOP occurrences over the total available resonances) in the wavelength range from $1540~{\rm nm}$ to $1561~{\rm nm}$ for $146~{\rm GHz}$ and $196~{\rm GHz}$ microresonators, measured at different pump power levels. 
  \textbf{b} Maximum on-chip SH power recorded at different pump power levels, for both microresonators during the wavelength sweeps as in \textbf{a}. 
  \textbf{c} Maximum on-chip conversion efficiency (CE) for different SH modes of the $146~{\rm GHz}$ microresonator, measured at $22.2~{\rm dBm}$ pump power. \textbf{d} Estimated internal CE (CE$_{\rm R}$) for different SH modes based on the measured CE and simulated external coupling rate at SH.
  All the measurements are performed without having external temperature control of the chips.}
 \label{Figure4n}
\end{figure*} 

Given the FSR difference between the fundamental pump mode and the SH mode (see Supplementary Note 1), the interaction of the same pump-SH mode pair can be matched several times within the pump sweep in the telecom-band (Figure \ref{Figure2n}a).
% In particular, we are interested in the pair between the fundamental pump mode and $4^{\rm th}$ SH mode (SH4) because of the large QPM grating period.
For a specific pump-SH mode pair, 
the mismatch of the participating azimuthal mode numbers $m_{\rm b}-2m_{\rm a}$ is however not identical throughout the sweep. 
Therefore, the number of $\chi^{(2)}$ grating period $N = |m_{\rm b}-2m_{\rm a}|$ inscribed inside the microresonator should vary accordingly. 
Using TP imaging, we confirm such changes at different pump resonances in the $146~{\rm GHz}$ microresonator.
In particular, we study the fundamental pump mode-SH4 pair because of the large resulting QPM period. We experimentally identify three such AOP occurrences, which take place at $1542.90~{\rm nm}$, $1549.10~{\rm nm}$ ($T= 45~{\rm ^\circ C}$) and $1559.35~{\rm nm}$. 
At these wavelengths, the inscribed $\chi^{(2)}$ grating structures are recorded in Figure \ref{Figure3n}e, with the extracted grating periods of $90.1~{\rm \mu m}$, $80.6~{\rm\mu m}$ and $70.7~{\rm\mu m}$, respectively. 
As shown in Figure \ref{Figure3n}f, the retrieved grating periods are in excellent agreement with simulated periods considering the fabrication tolerances. 
They roughly correspond to 11, 12, and 14 periods of QPM gratings in the microresonator, calculated based on $N= 2\pi R/\Lambda$.
The slight discrepancy from the integer numbers is likely due to the imprecision in grating period measurement.
Moreover, by superimposing a number of depth-scanned TP images, we are able to reconstruct the TP image of the entire microresonator poled at $1543.00~{\rm nm}$ (the same resonance as $1542.90~{\rm nm}$). 
11 QPM periods are clearly recognized in Figure \ref{Figure3n}g, matching the estimated number of periods above. 
Besides, the FSRs at fundamental pump mode and SH4 are simulated to be $145.7~{\rm GHz}$ and $131.9~{\rm GHz}$, respectively (see Supplementary Note 1). 
If we denote the participating azimuthal mode numbers at $1542.90~{\rm nm}$ as $(m_{\rm a}',m_{\rm b}')$, 
we could deduce from the simulated FSRs that, the azimuthal numbers involved at $1549.10~{\rm nm}$ and $1559.35~{\rm nm}$ are respectively $(m_{\rm a}'-5,m_{\rm b}'-11)$ and $(m_{\rm a}'-14,m_{\rm b}'-31)$.
The relative changes in the azimuthal number differences $m_{\rm b}- 2m_{\rm a}$ are also consistent with the retrieved QPM period numbers above. 

\vspace{0.1cm}

\noindent\textbf{Performance of SH generation in Si$_3$N$_4$ microresonators.}  
The performance of photo-induced SH generation in Si$_3$N$_4$ microresonators is illustrated in Figure \ref{Figure4n}. 
To initiate AOP, the pump and SH need to be doubly resonant. 
As the pump and SH resonances show distinct thermal shifts (see Supplementary Note 1), the initial mismatch between them can be effectively compensated during the thermal-induced resonance drag at pump.
Therefore, with higher pump power, 
the probability of AOP occurrence also increases 
with the extended thermal triangle. 
As shown in Figure. \ref{Figure4n}a, this is experimentally demonstrated in $146~{\rm GHz}$ and $196~{\rm GHz}$ Si$_3$N$_4$ microresonators. 
The AOP occurrences are recorded by varying the input pump power throughout the wavelength range between $1540~{\rm nm}$ to $1561~{\rm nm}$. 
Specifically, by pumping the $146~{\rm GHz}$ microresonator with $23.7~{\rm dBm}$ input power, the microresonator can be poled at $14$ out of $18$ available resonances. 
We also notice that higher AOP probability is achieved for the $146~{\rm GHz}$ microresonator than the $196~{\rm GHz}$ one.
This is mainly attributed to the smaller FSR and larger cross-section of the $146~{\rm GHz}$ microresonator (see Supplementary Note 1), which holds higher probability for resonance matching and supports more interacting modes at SH.
Figure \ref{Figure4n}b records the maximum generated SH power on-chip throughout the wavelength sweeps. 
It is evident that the maximum SH power grows with the input pump power, which is generally obtained at higher-order SH mode due to the large external coupling rate (see Supplementary Note 1). 
Moreover, both microresonators could output more than $10~{\rm dBm}$ of SH powers with moderate pump powers. 

The maximum on-chip CE for each SH mode, measured at the output of the $146~{\rm GHz}$ microresonator with $22.2~{\rm dBm}$ pump power is plotted in Fig. \ref{Figure4n}c. The highest CE is generated from the higher-order SH modes (4$^{\rm th}$ or 5$^{\rm th}$), dominated by the SH coupling rate. The coupling from the microresonator to the bus varies significantly between SH modes, and is extremely small for lower-order SH mode (see Supplementary Note 1). 
Figure \ref{Figure4n}d shows the estimated internal CE (${\rm CE_{R}}$) based on the measured CE and the simulated coupling at SH. Contrary to the output CE, the internal CE is maximized for the lower-order SH mode due to the minimal loss rate and optimized nonlinear overlap (see Supplementary Note 1). The highest estimated internal CE reaches  $3.0 \times 10^6~\%/{\rm W}$ for the fundamental SH mode, suggesting that the current output CE can be significantly improved with efficient output coupling. 

\section*{Discussion} 
 A comprehensive comparison of state-of-the-art microresonator platforms for SH generation is summarized in Table \ref{Table1}. 
 The ultrabroad bandwidth demonstrated in AOP of Si$_3$N$_4$ microresonators allows for tunable SH generation within a pump resonance, unparalleled by most of the other microresonators. 
%  In addition, we are able to obtain SH generation for 14 out of 18 available resonances in the $146~{\rm GHz}$ microresonator. 
Since our devices are not optimized for the out coupling at SH, the CE on-chip is limited when compared to $\chi^{(2)}$ microresonators based on LN\cite{chen2019ultra,lu2019periodically,lu2020toward} and AlN\cite{guo2016second,bruch201817}. 
However the internal CE indicates the full potential, with possible orders of magnitude improvements, as has been recently reported in ref\cite{lu2020efficient}. Still, our devices could output the highest SH power among all the platforms. 
Such high power and tunable SH generation is favored in many practical applications, for example the \textit{f-2f} self-referencing of Kerr combs on-chip\cite{miller2014chip,xue2017second}. 
Most importantly, the automatic QPM in AOP greatly facilitates the SH generation in Si$_3$N$_4$ microresonators. 
In the experiment, we could easily inscribe and reconfigure the self-organized $\chi^{(2)}$ gratings for nearly all the pumped resonances. This is in stark contrast with delicate intermodal phase-matching\cite{guo2016second,bruch201817,roland2016phase,zhang2019symmetry,lukin20204h} and $\overline{4}$-QPM\cite{kuo2014second,chang2019strong,mariani2014second,lake2016efficient,logan2018400}, or complicated poling achieved by high voltage sources\cite{wolf2018quasi,chen2019ultra,lu2019periodically,lu2020toward} required in other platforms.
Also as highlighted in the shaded part of Table \ref{Table1}, the recent progress made for SH generation in integrated silicon photonic platforms\cite{zhang2019symmetry,lukin20204h,levy2011harmonic,lu2020efficient} should not be overlooked. 
Undeniably, Si$_3$N$_4$ stands out as the most promising candidate in nearly every aspect of resonant SH generation.
Besides the recent demonstration of ultrahigh CE\cite{lu2020efficient}, 
we address in this work the unique benefit of  extremely flexible photo-induced QPM in Si$_3$N$_4$ microresonators.
These findings tend to indicate that past works on SH generation\cite{levy2011harmonic} (or SH combs\cite{miller2014chip,xue2017second}) in Si$_3$N$_4$ microresonators are most likely due to QPM instead of intermodal phase-matching, unless specifically designed for exact matching of azimuthal mode numbers. 

{\renewcommand{\arraystretch}{1.4}
\begin{table*}[hbt!]
\begin{threeparttable}
\caption{Comparison of state-of-the-art microresonator platforms for second-harmonic generation.
\label{Table1}
}
\begin{center}
\footnotesize\centering
\begin{tabular}{ m{2.6cm}<{\centering} m{2.8cm}<{\centering} m{2.8cm}<{\centering} m{3cm}<{\centering} m{4.9cm}<{\centering} }
\hline
\textbf{Platform} & \textbf{10 dB bandwidth} & \textbf{On-chip CE} & \textbf{On-chip SH power} & \textbf{Phase-matching (PM) condition}  \\
\hline
LN\cite{wolf2018quasi} & NA & $ 90\%/{\rm W}$ & $ 22.5~{\rm \mu W}$ & QPM (electric-field poling) \\
%LN\cite{lin2019broadband} & NA & $ 9900\%/{\rm W}$ & $ 25~{\rm \mu W}$ & natural QPM \\
LN\cite{chen2019ultra} & NA & $ 230000\%/{\rm W}$ & $ 2.1~{\rm \mu W}$ & QPM (electric-field poling)\\
LN\cite{lu2019periodically} & $ 1.9~{\rm pm}$ & $ 250000\%/{\rm W}$ & $ 0.1~{\rm mW}$ & QPM (electric-field poling)\\
LN\cite{lu2020toward} & $ 0.4~{\rm pm}$ & $ 5000000\%/{\rm W}$ & $ 20~{\rm \mu W}$  & QPM (electric-field poling) \\

AlN\cite{guo2016second} & $ 418~{\rm pm}$ & $ 2500\%/{\rm W}$ & $ 3.2~{\rm mW}$  & Intermodal PM \\
AlN\cite{bruch201817} & NA & $ 17000\%/{\rm W}$ & $ 10~{\rm mW}$  & Intermodal PM\\

GaN\cite{xiong2011integrated} & NA  & $ 0.015\%/{\rm W}$ & $ 2.2~{\rm \mu W}$ &  Intermodal PM (claimed)\\
GaN\cite{roland2016phase} & $ 200~{\rm pm}$ & $ 0.0002\%/{\rm W}$ & $ 2.4~{\rm pW}$ & Intermodal PM\\

GaAs\cite{kuo2014second} & $ 300~{\rm pm}$  & $ 5\%/{\rm W}$ & $ 0.6~{\rm nW}$ & $\overline{4}$-QPM \\
GaAs\cite{chang2019strong} & $ 20~{\rm pm}$ & $ 100\%/{\rm W}$ & $ 13~{\rm \mu W}$ & $\overline{4}$-QPM \\

AlGaAs\cite{mariani2014second}  & $ 2000~{\rm pm}$ & $ 0.07\%/{\rm W}$ & $ 5~{\rm nW}$  & $\overline{4}$-QPM \\

GaP\cite{lake2016efficient}  & NA & $ 44\%/{\rm W}$ & $ 47~{\rm nW}$ & $\overline{4}$-QPM \\

GaP\cite{logan2018400}  & NA & $ 400\%/{\rm W}$ & $ 25~{\rm \mu W}$ & $\overline{4}$-QPM \\

\rowcolor{mygray}
SiO$_2$\cite{zhang2019symmetry} & $ 13~{\rm pm}$  & $ 0.049\%/{\rm W}$ & $ 6~{\rm nW}$  & Intermodal PM\\

\rowcolor{mygray}
SiC\cite{lukin20204h}  & NA & $ 360\%/{\rm W}$ & $ 1~{\rm \mu W}$ & Intermodal PM\\

\rowcolor{mygray}
Si$_3$N$_4$\cite{levy2011harmonic} & NA  & $ 0.1\%/{\rm W}$ & $ 0.1~{\rm mW}$  & Intermodal PM (claimed)\\
\rowcolor{mygray}
Si$_3$N$_4$\cite{lu2020efficient} &  $ 60~{\rm pm}$  & $ 2500\%/{\rm W}$ & $ 2.2~{\rm mW}$  & Intermodal PM\\ 
\rowcolor{mygray}
Si$_3$N$_4$ (this work) & \textbf{$ 605~{\rm pm}^*$}  & $  51\%/{\rm W}$ & $  12.5~{\rm mW}$ & Reconfigurable QPM\\

\hline
\end{tabular}
\begin{tablenotes}
\item 
NA: data not available. \hspace{5mm} The best values extracted from references are shown in the table. \\
Shaded area refers to typical silicon photonic platforms.\\
$^*$Bandwidth for a single pump resonance. SH generation is achieved for 14 out of 18 resonances in telecom-band.
\end{tablenotes}
\end{center}
\end{threeparttable}
\end{table*}
}

We explore in this study AOP of $146~{\rm GHz}$ and $196~{\rm GHz}$ Si$_3$N$_4$ microresonators for SH generation. 
Leveraging small FSR and overmoded microresonators, AOP is easily induced in multiple pumped resonances.
Temperature and pump power controls offer additional degrees of freedom for optimizing the doubly resonant condition.
While the AOP threshold could be lowered and efficiency increased by using microresonators with larger FSRs\cite{lu2020efficient}, the possibility of AOP is significantly reduced.
Therefore, given the recent development of Si$_3$N$_4$ microresonators towards high Qs and small FSRs\cite{puckett2021422,2020arXiv200513949L}, we clearly envision SH generation with combined ultrahigh CE and poling probability in small FSRs designs. Under such conditions, SH may be generated for every single resonance in an ultrafine FSR grid, thanks to automatic QPM rather than intermodal phase-matching. 
From various VNA measurements, we also note that AOP could take place prior to the exact detuning condition mediated by the integer number of grating periods, i.e. $\theta_{\rm b} =2\theta_{\rm a}$ or $\delta_{\rm b} \approx 2\delta_{\rm a}$. 
We even observe AOP when the fundamental SH resonance is at the blue side of the generated SH ($\delta_{\rm b}>0$ and $\delta_{\rm a}<0$), as seen in Supplementary Note 3.
These observations provide important insights on the detuning information at AOP initiation, being slightly in advance of the exact QPM condition.

In conclusion, we have demonstrated versatile SH generation in Si$_3$N$_4$ microresonators via photo-induced $\chi^{(2)}$ gratings. 
The grating structures are organized following the interference of doubly resonant pump and SH fields, which are extremely flexible and automatically adapt to the required QPM periods. 
We experimentally characterized the inscribed $\chi^{(2)}$ gratings by TP imaging, as well as evaluated the induced $\chi^{(2)}$ strengths. 
We also implement a method for tracking the interacting resonance detunings during poling events, and observed ultrabroad SH generation bandwidth in AOP process. Such technique, based on VNA measurement, is a powerful way to study all kinds of resonant harmonic generation.
The findings in this work not only further anchor $\chi^{(2)}$ nonlinearity in the toolbox of silicon photonics, 
but also motivate a SH generation paradigm with fully reconfigurable QPM mechanism and simultaneous high performance (in terms of CE, SH power, bandwidth, etc). 
Similar to Pockels materials\cite{he2019self,bruch2021pockels}, the combination of $\chi^{(2)}$ and $\chi^{(3)}$ nonlinearities in Si$_3$N$_4$ microresonators would trigger new nonlinear applications in CMOS-compatible platforms.

\vspace{0.5cm}

\noindent\textbf{Methods}
\medskip
\begin{footnotesize}
%\begin{methods}

 \noindent\textbf{Optical poling and detuning measurement setups}:
 A tunable telecom-band CW laser is polarization controlled, and amplified with an erbium-doped fiber amplifier (EDFA). The amplified pump is aligned at TE polarization, and injected to the bus waveguide of a Si$_3$N$_4$ microresonator using a lensed fiber. The input coupling loss is estimated to be $2.4~{\rm dB}$.
 At the output of the chip, both the residual pump and the generated SH are collected using a microscope objective before being separated by a dichroic beamsplitter and directed to their respective photodetectors (PD1 and PD2). 
 For measuring the effective detunings of both pump and SH resonances, the CW laser is first weakly phase-modulated in a EOM before amplification. 
 The applied modulation signal is a sweeping microwave tone from the VNA which scans from $5~{\rm kHz}$ to $1.5~{\rm GHz}$. 
 Two weak optical sidebands are thus created at the pump wavelength, amplified together with the pump and coupled to the microresonator. 
 Part of the generated light at SH is tapped to a reflective collimator and then directed a AC-coupled fast silicon photodector (PD3) with $1 ~{\rm GHz}$ bandwidth. 
 Finally, the retrieved microwave signal is sent back to the VNA. 
 It is noted that although our probing method resembles ref.\cite{guo2017universal}, the fundamental difference is the use of the photodetector at SH wavelength rather than at pump wavelength. 
 Under such condition, no signal is detected before AOP occurs. 
 When the $\chi^{(2)}$ is inscribed, in addition to SH generation, there is also sum-frequency generation between the sidebands and pump. Eventually, the beating between the sum-frequency components and the SH component gives the VNA response. 
 The transfer function in VNA is found to reflect detuning information of both pump and SH resonances (see Supplementary Note 2). 

\vspace{0.1cm}

\noindent\textbf{$\chi^{(2)}$ grating imaging and estimation}: 
The inscribed $\chi^{(2)}$ gratings in poled microresonators are measured with a TP microscope (Leica LSM $710$ NLO) in an upright configuration. 
%at the Cellular Imaging Facility of University of Lausanne. 
A He-Ne laser operating at $633 ~{\rm nm}$ is employed for microscope alignment. 
For excitation of SH in poled microresonators, we use a Ti:Sapphire laser (Coherent Chameleon Ultra II IR) producing $1010 ~{\rm nm}$ horizontally polarized light relative to the image plane. 
In the measurement, the focal point is raster-scanned across the sample in the grating plane, and then its generated SH signal is recorded. 
The measured grating shape is slightly distorted in regions where the waveguide tangent is not perpendicular to the incident light polarization.
The optical resolution of the obtained TP images is estimated to be $310 ~{\rm nm}$. 
These images are then replotted in the radial and tangential coordinates of the microresonators (see Supplementary Note 4), as shown in Figure \ref{Figure2n}c. As such, the $\chi^{(2)}$ grating periods can be easily extracted and their spatial frequencies are shown in Figure \ref{Figure2n}d. 
The $\chi^{(2)}$ values in  poled microresonators are estimated based on the comparison with calibrated $\chi^{(2)}$ in a poled waveguide (see Supplementary Note 5). 

\vspace{0.1cm}

\noindent \textbf{Acknowledgements}:
This work was supported by ERC grant PISSARRO (ERC-2017-CoG 771647). We thank Ozan Yakar for assistance in poling and characterization of straight waveguides, and Wenle Weng for helpful discussions. 

\vspace{0.1cm}

\noindent \textbf{Data Availability Statement}: 
The data and code that support the plots within this paper and other findings of this study are available from the corresponding authors upon reasonable request.
\end{footnotesize}

\bibliographystyle{naturemag}
\bibliography{ref}

\end{document}

\title{Supplementary Information to: \\ Optically reconfigurable quasi-phase-matching in silicon nitride microresonators}

\author{Edgars Nitiss}
\thanks{These authors contributed equally to the work.}
\affiliation{{\'E}cole Polytechnique F{\'e}d{\'e}rale de Lausanne, Photonic Systems Laboratory (PHOSL), STI-IEL, Station 11, Lausanne CH-1015, Switzerland.}
 
\author{Jianqi Hu}
\thanks{These authors contributed equally to the work.}
\affiliation{{\'E}cole Polytechnique F{\'e}d{\'e}rale de Lausanne, Photonic Systems Laboratory (PHOSL), STI-IEL, Station 11, Lausanne CH-1015, Switzerland.}

\author{Anton Stroganov}
\affiliation{LIGENTEC SA, EPFL Innovation Park, Bâtiment L, 1024 Ecublens, Switzerland.}

\author{Camille-Sophie Br\`es}
\email[]{camille.bres@epfl.ch}
\affiliation{{\'E}cole Polytechnique F{\'e}d{\'e}rale de Lausanne, Photonic Systems Laboratory (PHOSL), STI-IEL, Station 11, Lausanne CH-1015, Switzerland.}

\maketitle

%%%%%%%%%%%%%%%%%%%%%
\section*{\textbf{Supplementary Note 1. Si$_3$N$_4$ microresonator devices}}

% FIGURE WITH CHIP PARAMETERS
\begin{figure}[htp]
  \renewcommand{\figurename}{Supplementary Figure}
  \centering{
  \includegraphics[width = 0.9\linewidth]{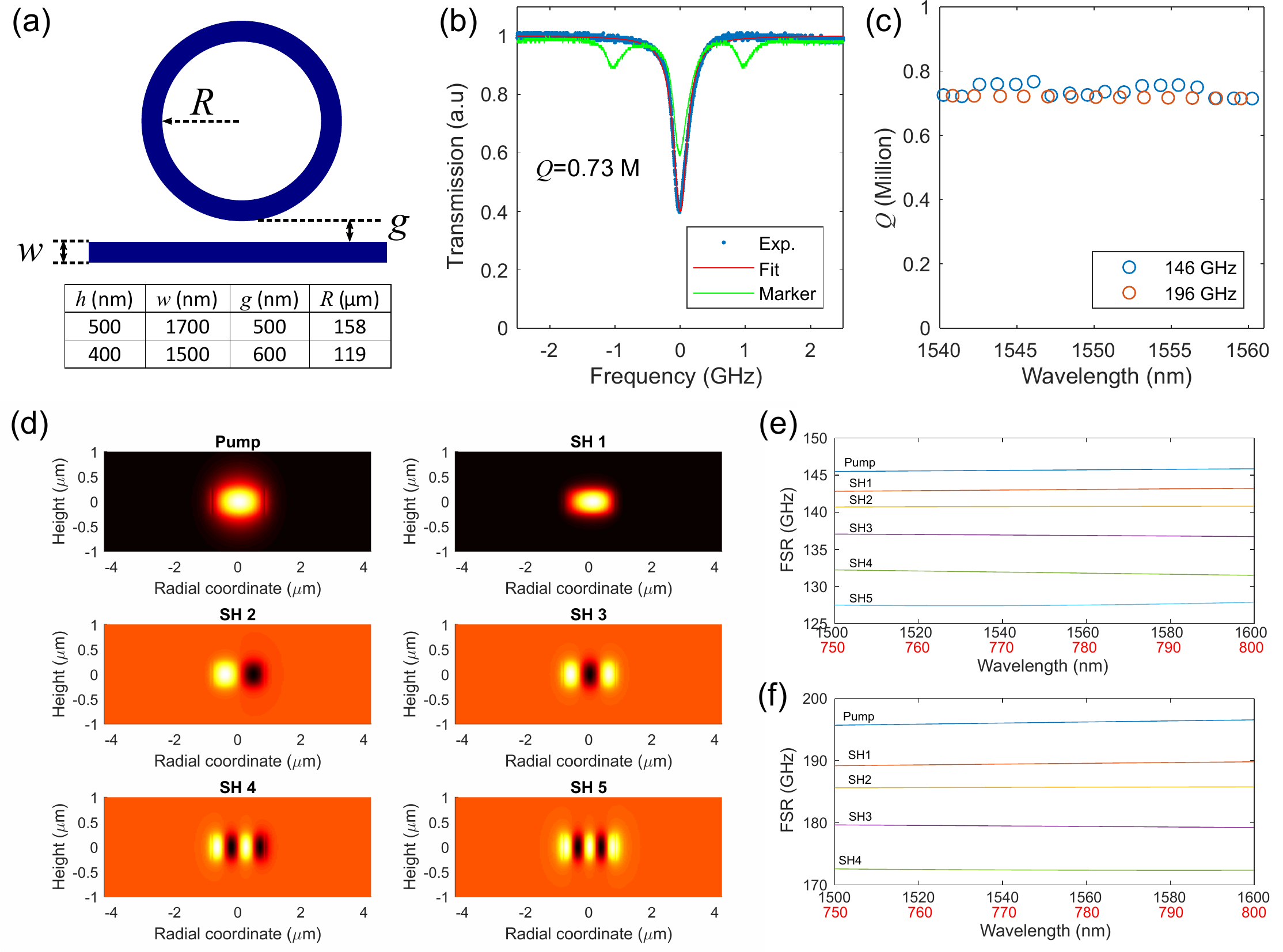}
  } 
  \caption{\noindent\textbf{The design, simulation, and characterization of microresonator devices.} \textbf{a} Schematic of the Si$_3$N$_4$ microresonator device comprising a ring and a bus waveguide. The table holds parameters of microresonators used in the study. h: height; w: width; g: gap width; R: ring radius. The ring and bus waveguides are designed to have identical cross-sections. 
  \textbf{b} Resonance linewidth measurement of the ${146~{\rm GHz}}$ microresonator at resonance of $1542.57 ~{\rm nm}$. The measured resonance linewidth (blue) is calibrated by the marker trace (green) being modulated by $1~{\rm GHz}$ signal, and fitted with a Lorentzian function (red). The loaded quality factor (Q) is measured to be around $0.73 \times 10^{6}$. 
  \textbf{c} Measured loaded Qs of both ${146~{\rm GHz}}$ and ${196~{\rm GHz}}$ microresonators in the wavelength range from ${1540~{\rm nm}}$ to ${1561~{\rm nm}}$. 
  \textbf{d} TE-polarized mode field distributions in the ${146~{\rm GHz}}$ microresonator. The fundamental mode at pump wavelength (Pump) and the first five modes at SH wavelength (SH1-5) are shown. 
  \textbf{e} and \textbf{f} FSRs of fundamental mode at pump (Pump) and various modes at SH (SH1-5) in microresonators with ring radii $R= 158~{\rm \mu m}$ and $R= 119~{\rm \mu m}$, respectively, as a function of wavelength. Here pump FSRs are plotted for wavelength range from $1500~{\rm nm}$ to $1600 ~{\rm nm}$, while from $750 ~{\rm nm}$ to $800 ~{\rm nm}$ for SH. 
  }
 \label{FigureS1}
\end{figure} 

In the study we employed Si$_3$N$_4$ microresonators with ring radii of $119~{\rm \mu m}$ and $158~{\rm \mu m}$, fabricated by LIGENTEC using its AN-technology platform. In the C-band, the free spectral range (FSR)  approximately corresponds to $146~\rm {GHz}$ and $196~\rm {GHz}$, respectively.  
The ring structures are coupled with straight waveguides of identical cross-sections, with all parameters indicated in Supplementary Figure \ref{FigureS1}a. 
At the input and output of the bus waveguides, tapers are used for efficient light coupling for a pump wavelength in the telecommunication band. 
The resonance linewidths of the microresonator devices are characterized by the sideband modulation technique\cite{li2012sideband}. 
An example of recorded data is shown in Supplementary Figure \ref{FigureS1}b. By scanning a continuous-wave (CW) laser over the resonance at $1542.58~\rm {nm}$ of the $146~\rm {GHz}$ microresonator, the transmission data (blue dots) are recorded and then fitted with a Lorentzian lineshape (red line). 
In order to calibrate the resonance linewidths, the laser was modulated by a $1~{\rm GHz}$ tone to create two sidebands, with each sideband separated $1~{\rm GHz}$ away from the optical carrier. 
In this scenario, two additional resonances would appear in the measured transmission trace (green line), serving as frequency markers.
Supplementary Figure \ref{FigureS1}c shows the loaded quality factor (Q) values of all resonances in the wavelength range from $1540 ~{\rm nm}$ to $1561 ~{\rm nm}$ for both microresonators. 
The loaded Q values in both devices are measured to be around $0.75\times 10^6$.

{\renewcommand{\arraystretch}{1.4}
\begin{table}[htp]
  \renewcommand{\tablename}{Supplementary Table}
    \caption{\textbf{Parameters for fundamental mode at pump and various modes at SH of 146 GHz and 196 GHz microresonators.} 
    The effective refractive indices $n_{\rm eff}$, effective and material thermo-optic coefficients $dn_{\rm eff}/dT$, $dn_{\rm Si3N4}/dT$, $dn_{\rm SiO2}/dT$, temperature dependent wavelength shifts $d\lambda/dT$, nonlinear spatial overlaps $\Gamma$ as well as the external coupling rates $\kappa_{\rm ex}$ for fundamental TE mode at pump and lower-order TE modes at SH, calculated or measured at $1550 ~{\rm nm}$ for pump and at $775~{\rm nm}$ for SH.
    }
    \label{S_table1}
    \begin{center}
    \footnotesize\centering
    %% TABLE FOR 150 GHZ RING
    \begin{tabular}{ m{4.0cm}<{\centering} m{1.2cm}<{\centering} m{1.2cm}<{\centering} m{1.2cm}<{\centering} m{1.2cm}<{\centering} m{1.2cm}<{\centering} m{1.2cm}<{\centering}}
    \multicolumn{7}{m{11.2cm}<{\centering}}{\small $146~{\rm GHz}$ microresonator, width $\times$ height: $1.7 \times 0.5~{\rm \mu m^2}$, gap: $500~{\rm nm}$}\\
    \hline
    \textbf{Mode} & \textbf{FH1} & \textbf{SH1} & \textbf{SH2} & \textbf{SH3} & \textbf{SH4} & \textbf{SH5}  \\ 
    \hline
    $n_{\rm eff}$ & $1.742$ & $1.918$ & $1.881$ & $1.819$ & $1.731$ & $1.619$  \\
    $dn_{\rm eff}/dT~(10^{-6}/{\rm K})$  &$21.9$ & $33.0$  & $33.3$ & $34.0$ & $34.8$ & $34.9$  \\
    $dn_{\rm Si3N4}/dT~(10^{-6}/{\rm K})$  &$22.8$ & $32.7$  & $32.7$ & $32.7$ & $32.7$ & $32.7$  \\
    $dn_{\rm SiO2}/dT~(10^{-6}/{\rm K})$  &$8.54$ & $8.70$  & $8.70$ & $8.70$ & $8.70$ & $8.70$  \\
    $d\lambda/dT~({\rm pm/K})$  &$19.49$ & $13.32$  & $13.74$ & $14.49$ & $15.58$ & $16.71$  \\
    $\kappa_{\rm ex}/2\pi~({\rm MHz}) $ & %$139$ 
    & $0.04$ & $0.27$ & $2.32$ & $28.0$ & $433$  \\
    $\Gamma~({\rm \%})$ & & $97.5$ & $5.7$ & $13.3$ & $0.4$ & $1.2$  \\
    
    %% TABLE FOR 200 GHZ RING
    \hline\\
    \end{tabular}
    \begin{tabular}{ m{4.0cm}<{\centering} m{1.2cm}<{\centering} m{1.2cm}<{\centering} m{1.2cm}<{\centering} m{1.2cm}<{\centering} m{1.2cm}<{\centering}}
    \multicolumn{6}{m{10cm}<{\centering}}{\small $196~{\rm GHz}$ microresonator, width $\times$ height: $1.5 \times 0.4~{\rm \mu m^2}$, gap: $600~{\rm nm}$} \\
    \hline
    \textbf{Mode} & \textbf{FH1} & \textbf{SH1} & \textbf{SH2} & \textbf{SH3} & \textbf{SH4} \\ 
    \hline
    $n_{\rm eff}$ & $1.680$ & $1.884$ & $1.837$ & $1.757$ & $1.646$ \\
    $dn_{\rm eff}/dT~(10^{-6}/{\rm K})$  &$20.7$ & $32.7$  & $33.1$ & $33.8$ & $34.3$ \\

    $d\lambda/dT~({\rm pm/K})$  &$19.10$ & $13.44$  & $13.98$ & $14.92$ & $16.14$ \\
    $\kappa_{\rm ex}/2\pi~({\rm MHz}) $ & %$242$ 
    & $0.02$ & $0.13$ & $2.08$ & $57.7$ \\
    $\Gamma~({\rm \%})$ & & $96.5$ & $6.0$ & $11.3$ & $0.07$\\
    
    \hline\\
    \end{tabular}
    \end{center}
\end{table}
}

We use COMSOL Multiphysics to simulate the effective refractive indices $n_{\rm eff}$ of the interacting TE-polarized pump and second-harmonic (SH) modes, as well as their mode field distributions. 
Supplementary Figure \ref{FigureS1}d illustrates the different mode shapes of a $146~{\rm GHz}$ microresonator at pump wavelength around $1550~{\rm nm}$ and SH wavelength around $775~{\rm nm}$. 
The spatial mode distributions were then utilized to simulate the structure of the inscribed grating inside the microresonator. 
In addition, they were used to calculate the nonlinear spatial overlap $\Gamma$ between the fundamental mode and different SH modes (see Supplementary Table \ref{S_table1}) as defined in ref\cite{lu2020efficient}. 
The FSRs of microresonators are extracted based on the simulated $n_{\rm eff}$ wavelength dependence.
For both microresonators, the calculated FSRs of fundamental mode at pump and various modes at SH are depicted in Supplementary Figure \ref{FigureS1}e and f, respectively. 
We also account for the external coupling rates $\kappa_{\rm ex}$ of the SH modes (see Supplementary Table \ref{S_table1}) based on finite-difference time domain (FDTD) simulation\cite{pfeiffer2017coupling}. 

\begin{figure}[htp]
  \renewcommand{\figurename}{Supplementary Figure}
  \centering{
  \includegraphics[width = 0.7\linewidth]{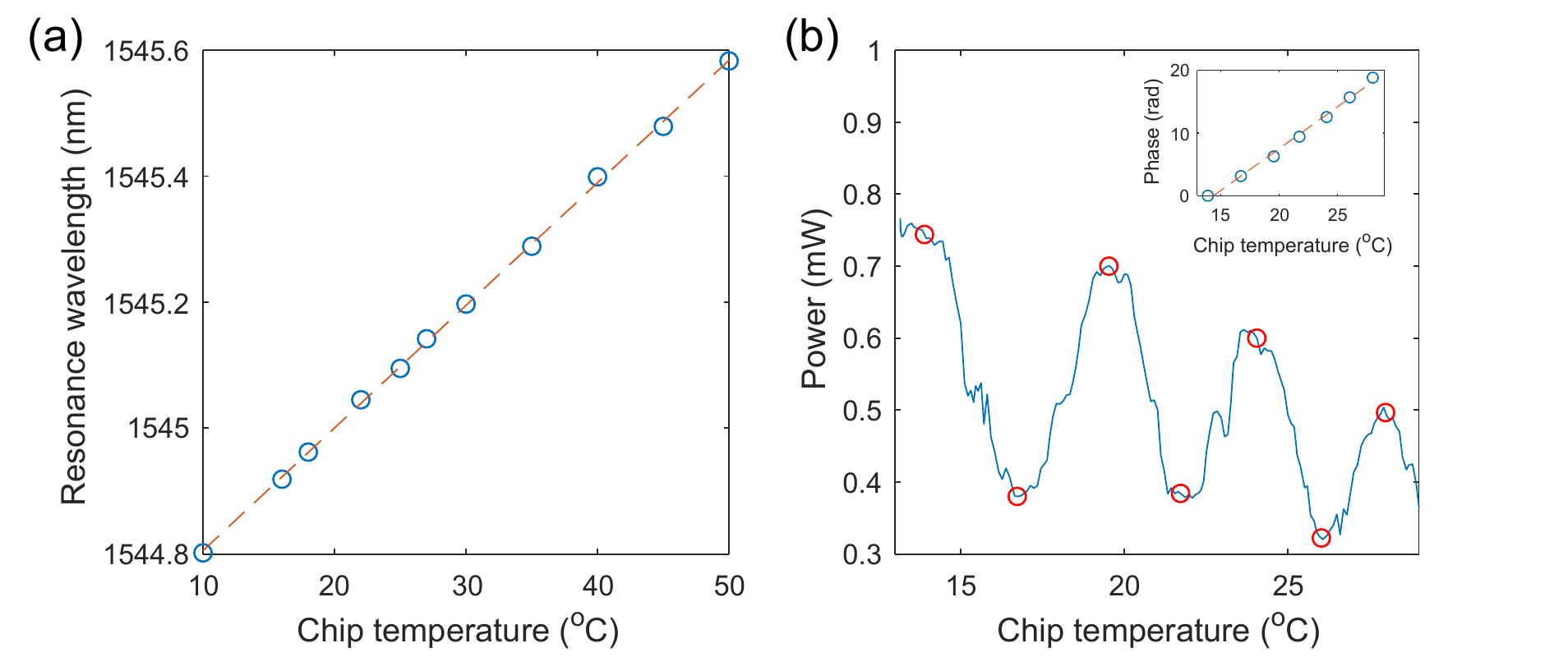}
  } 
  \caption{\noindent\textbf{Measurements of effective thermo-optic coefficients in the 146 GHz Si$_3$N$_4$ microresonator. a} Pump resonance wavelength as a function of sample temperature. Blue: experimental data; Dashed orange: linear fit. $d\lambda/dT=19.49~{\rm pm/K}$ is extracted which gives the effective thermo-optic coefficient at pump as $dn_{\rm eff}/dT=21.9\times 10^{-6}~{\rm K}^{-1}$. \textbf{b} Interferogram recorded at the output of a Mach-Zehnder interferometer (MZI)  comprising a $775 ~{\rm nm}$ light source and a straight waveguide with temperature control in one of its arms. Blue: experimental data; Red: the estimated maximum and minimum points corresponding to the constructive or destructive interference, respectively. 
  Inset shows the extracted phase $\psi$ (blue) and its linear fitting (dashed orange) as a function of temperature. $d\psi/dT=1.34~{\rm rad/K}$ is estimated which gives $dn_{\rm eff}/dT=33.0\times 10^{-6}~{\rm K}^{-1}$ for fundamental mode at SH wavelength.
  } 
 \label{FigureS2}
\end{figure}

For the demonstrated all-optical poling (AOP) process, the doubly resonant condition for both pump and SH is met thanks to the differential thermo-optic and Kerr resonance shift at their respective wavelengths. 
The temperature change in microresonators, which happens whenever pumping a resonance as well as under different chip temperature tuning, enables compensating for an initial cavity resonance mismatch as to initiate AOP. 
% Therefore knowledge of the thermo-optic coefficients is essential in the device design step. 
The thermo-optic coefficients of the microresonators are thus important and estimated by several experiments and simulations.
The temperature dependent wavelength shift at pump wavelength was determined by measuring the pump resonance position as a function of sample temperature (Supplementary Figure \ref{FigureS2}a). 
We extract $d\lambda/dT=19.49~{\rm pm/K}$ at the pump wavelength. 
The effective thermo-optic coefficient $dn_{\rm eff}/dT$ was then calculated by:
\begin{equation}
\frac{d n_{\rm eff}}{d T} = \frac{d \lambda}{d T}\frac{ n_{\rm eff}}{\lambda}
\label{S1}
\end{equation}
where $\lambda$ is the resonance wavelength. From Eq. \eqref{S1} the $dn_{\rm eff}/dT$ is found to be $21.9\times 10^{-6} ~{\rm K}^{-1}$ at pump wavelength. 
We then estimate the thermo-optic coefficient $dn_{\rm Si3N4}/dT$ of Si$_3$N$_4$ based on COMSOL Multiphysics simulation, by assuming the uniform temperature distribution and using the thermo-optic coefficient of silica $dn_{\rm SiO2}/dT$ from ref\cite{leviton2006temperature}. 
The estimated thermo-optic coefficient of Si$_3$N$_4$ at pump wavelength $dn_{\rm Si3N4}/dT=22.8\times 10^{-6}~{\rm K}^{-1}$ is in good agreement with the value reported previously\cite{arbabi2013measurements}. 
% FIGURE WITH RESONANCE THERMAL SHIFTS

Since the device is significantly under-coupled for the fundamental mode at SH (see Supplementary Table \ref{S_table1}), we were not able to conduct the same experiment for the estimation of thermo-optic coefficients at SH band. 
In an alternative approach to determine the effective thermo-optic coefficients at SH wavelength, we measured the phase difference induced in the bus waveguide as a function of chip temperature in a Mach-Zehnder interferometer (MZI).
Here a CW laser operating at $775~{\rm nm}$ was split into two arms with the sample inserted in one of the arms. 
At the output of the MZI, the two beams were recombined and their resulting interference pattern at different chip holder temperatures is recorded (see Supplementary Figure \ref{FigureS2}b). 
We then extracted the induced phase shift with respect to temperature $d\psi/dT$ from temperature values at which constructive and destructive interference took place. 
From a linear fit we estimated the $d\psi/dT=1.34~{\rm rad/K}$. As such, the effective thermo-optic coefficient $dn_{\rm eff}/dT$  at SH wavelength can be obtained from:
\begin{equation}
\frac{d n_{\rm eff}}{d T}
= \frac{d \psi}{d T}\frac{\lambda}{2\pi L}
\label{S2}
\end{equation}
where $L=5~{\rm mm}$ is the bus waveguide length. $dn_{\rm eff}/dT=33.0\times 10^{-6}~{\rm K}^{-1}$ is thus estimated for fundamental mode at SH wavelength. 
Following the same approach as described above, the thermo-optic coefficient of Si$_3$N$_4$ at SH wavelength was extracted from its effective thermo-optic coefficient. 
While the effective thermo-optic coefficients for higher-order SH modes were also determined by COMSOL simulation, which eventually give their temperature dependent resonance wavelength shifts $dn_{\rm eff}/dT$ based on Eq. \eqref{S1}. 
All of the thermo-optic parameters for both microresonators used in this study are included in Supplementary Table \ref{S_table1}. 

As evident from Supplementary Table \ref{S_table1}, we found  $d\lambda/dT=13.32~{\rm pm/K}$ for the fundamental mode at SH in our $146~{\rm GHz}$ Si$_3$N$_4$ microresonator, which is in a good agreement with the direct resonance frequency measurement reported in ref\cite{sinclair20201}. 
And similar to ref\cite{lu2020efficient}, for our devices the resonance shifts $d\lambda/dT$ of SH modes are larger than half of the resonance shifts at pump. 
Thus, when the pump is tuned on resonance so that the chip heats up, the differential resonance frequency shift is responsible for compensating the initial cavity resonance mismatch. 
Once it becomes doubly resonant, quasi-phase-matching (QPM) occurs and a grating is inscribed inside the microresonator. 
Noticeably, after the perfect resonance matching point, the SH mode resonance would also gradually walk off from the SH frequency when further detuning the pump. The effective detuning of SH mode is monitored by the vector network analyzer (VNA) measurement (see Supplementary Note 2).  
% It is important to note that the difference in thermo-optic resonance shifts is quite large – close to $3~{\rm pm/K}$ at SH wavelength. This suggests that a perfect resonance matching should occur only in a very narrow spectral range. Slight variation of poling wavelength in the resonance matched case should result in offset of cavity resonances due to change of resonator temperature.

\section*{\textbf{Supplementary Note 2. Operation principle of VNA measurement}}

In the main text, we presented a novel method of probing both the pump and SH cavity resonances via VNA measurement. 
The operation principle is described in this note. 
First, a CW pump laser is weakly phase modulated with a scanning RF tone $\omega_{\rm m}$ from a VNA. 
The modulated pump is then sent to microresonators for SH generation, and the generated SH signal is converted to electrical domain at a fast silicon photodetector. 
Note that in our approach, the cavity resonances are monitored with a photodetector operated at the SH wavelength, rather than at the pump wavelength as generally employed for microresonator detuning  monitoring\cite{guo2017universal}, as to allow for for the simultaneous extraction of both pump and SH detuning information.

It is well-known that weak phase modulation can be approximated as:
\begin{equation}
e^{i\epsilon\sin \omega_{\rm m} t} = 
\cos(\epsilon\sin \omega_{\rm m} t) + 
i\sin(\epsilon\sin \omega_{\rm m} t) \approx 1 + i\epsilon\sin \omega_{\rm m} t
= 1 + \frac{\epsilon}{2} {e^{i\omega_{\rm m}t}} - \frac{\epsilon}{2} {e^{-i\omega_{\rm m}t}}\label{S3}
\end{equation}
where $\epsilon$ is the strength of phase modulation. Under small signal approximation ($\epsilon \ll 1$), the modulation creates two weak sidebands at $\pm \omega_{\rm m}$ with respect to the pump frequency $\omega_{\rm p}$. Here we use the sign convention %of $e^{-i\omega t}$,
that $e^{\pm i\omega_{\rm m} t}$ represent the sideband frequencies of $\omega_p \mp \omega_m$. 
Once the SH is generated and inscribes the nonlinear $\chi^{(2)}$ grating in the microresonator, the sum-frequency process between the pump and sidebands will also take place, i.e. $\omega_{\rm p} + (\omega_{\rm p} \pm \omega_{\rm m}) \rightarrow 2\omega_{\rm p} \pm \omega_{\rm m}$, forming sidebands around the SH wavelength. The beating signal between the sum-frequency components and SH frequency gives the measured response in the VNA. 
The modulated pump and SH fields inside the microresonator can be described by coupled equations with the $\chi^{(2)}$ process as the coupling term\cite{leo2016frequency,weng2020heteronuclear}:
%\begin{equation}
\begin{gather}
\frac{\partial A}{\partial t}  = -\big(\frac{\kappa_{\rm a}}{2} + i\delta_{\rm a}\big)A + i gA^{*}B + \sqrt{\kappa_{\rm ex,a}}s_{\rm in} \big(1 + \frac{\epsilon}{2}e^{i\omega_{\rm m}t} -\frac{\epsilon}{2}e^{-i\omega_{\rm m}t} \big) 
\label{S4} \\[1ex] %
% \end{equation}
% \begin{equation}
\frac{\partial B}{\partial t} = -\big(\frac{\kappa_{\rm b}}{2} + i\delta_{\rm b}\big)B + i gA^2
\label{S5}
\end{gather}
where $A$ and $B$ are respectively the intracavity temporal envelopes of the pump and SH fields, normalized such that $|A|^2$ and $|B|^2$ represent the photon number. $\kappa_{\rm a (b)}$ are the total loss rates of pump and SH resonances, while $\kappa_{\rm ex,a}$ is the coupling strength of pump resonance. 
$\delta_{\rm a} = \omega_{\rm a} - \omega_{\rm p} $ and $\delta_{\rm b} = \omega_{\rm b} - 2\omega_{\rm p}$  are respectively the effective detunings at the pump and SH, while both resonance frequencies $\omega_{\rm a}$ and $\omega_{\rm b}$ take into account the effect of thermal and Kerr shifts. $g$ is the SH coupling strength, $|s_{\rm in}|^2 = P_{\rm p}/\hbar\omega_{\rm p} $ is the driving photon flux with $P_{\rm p}$ the input pump power and $\hbar$ the reduced Planck's constant. $\omega_{\rm m}$ is the VNA modulation frequency and $e^{\pm i\omega_{\rm m}t}$ denote the modulated sidebands, with $\epsilon/2$ representing their amplitudes normalized to the pump.  

In order to attain the VNA response, we first derive the steady-state solutions of Eq. \eqref{S4} and Eq. \eqref{S5} in the absence of pump modulation ($\epsilon = 0$). We consider in the approximation of no pump depletion, so that $igA^{*}B$ in Eq. \eqref{S4} can be omitted. The steady-state pump solution $A_0$ is thus obtained by setting $\frac{\partial A}{\partial t}=0$: 
\begin{equation}
A_0 = \frac{ \sqrt{\kappa_{\rm ex,a}}s_{\rm in}}{\frac{\kappa_{\rm a}}{2} + i\delta_{\rm a}}
\label{S6}
\end{equation}
Similarly, the steady-state SH solution $B_0$ can be computed by setting $\frac{\partial B}{\partial t}=0$ and substituting Eq. \eqref{S6} into Eq. \eqref{S5}: 
\begin{equation}
B_0 = \frac{igA_0^2}{\frac{\kappa_{\rm b}}{2} + i\delta_{\rm b}}
= \frac{ig \kappa_{\rm ex,a} s_{\rm in}^2}{(\frac{\kappa_{\rm a}}{2} + i\delta_{\rm a})^2(\frac{\kappa_{\rm b}}{2} + i\delta_{\rm b})}
\label{S7}
\end{equation}
Thus the out-coupled SH power is given by: 
\begin{equation}
P_{\rm SH} = \kappa_{\rm ex,b}\hbar 2\omega_{\rm p} |B_0|^2 
=  \frac{2 g^2 \kappa_{\rm ex,a}^2\kappa_{\rm ex,b}}{((\frac{\kappa_{\rm a}}{2})^2 + \delta_{\rm a}^2)^2
((\frac{\kappa_{\rm b}}{2})^2 + \delta_{\rm b}^2)}
\frac{P_{\rm p}^2}{\hbar \omega_{\rm p}}
\label{S8}
\end{equation}
where $\kappa_{\rm ex,b}$ is the coupling strength of the SH mode.

Now we consider the case when the pump is weakly modulated, i.e. the driving term becomes $s_{\rm in}(1+ \frac{\epsilon}{2}e^{i \omega_{\rm m }t} - \frac{\epsilon}{2}e^{-i \omega_{\rm m }t})$. Such modulation would also induce the perturbation of the intracavity pump field with an ansatz $A = A_0 + \epsilon A_{\rm -}e^{i\omega_{\rm m}t} + \epsilon A_{\rm +}e^{-i\omega_{\rm m}t}$. Inserting this ansatz back into Eq. \eqref{S4} derives the strengths of sidebands being coupled into the microresonator:
\begin{equation}
A_{\rm \mp} = \pm \frac{\sqrt{\kappa_{\rm ex,a}}s_{\rm in}}{2(\frac{\kappa_{\rm a}}{2}+ i\delta_{\rm a} \pm i\omega_{\rm m})}
\label{S9}
\end{equation}
We can see that the in-coupled sideband strengths are functions of the modulation frequency $\omega_{\rm m}$.
Such modulation at the pump also perturbs the SH field through the nonlinear coupling term $igA^2 = ig( A_0 + \epsilon A_{\rm -}e^{i\omega_{\rm m}t} + \epsilon A_{\rm +}e^{-i\omega_{\rm m}t})^2\approx
ig (A_0^2 + 2\epsilon A_0 A_{\rm -}e^{i\omega_{\rm m}t} + 2\epsilon A_0 A_{\rm +}e^{-i\omega_{\rm m}t}) $, where the higher-order terms are neglected ($\epsilon \ll 1$). 
Analogous to the derivation of $A_{\rm \mp}$, the modulation strengths $B_{\rm \mp}$ at SH can be obtained by inserting the ansatz $B = B_0 + \epsilon B_{\rm -}e^{i\omega_{\rm m}t} + \epsilon B_{\rm +}e^{-i\omega_{\rm m}t}$ into Eq. \eqref{S5}:
\begin{equation}
B_{\rm \mp} = \frac{2igA_0A_{\rm \mp}}{\frac{\kappa_{\rm b}}{2}+ i\delta_{\rm b} \pm i\omega_{\rm m}}
= \pm \frac{ig\kappa_{\rm ex,a}s_{\rm in}^2}{(\frac{\kappa_{\rm a}}{2}+ i\delta_{\rm a})(\frac{\kappa_{\rm a}}{2}+ i\delta_{\rm a} \pm i\omega_{\rm m})(\frac{\kappa_{\rm b}}{2}+ i\delta_{\rm b} \pm i\omega_{\rm m})}
\label{S10}
\end{equation}
$B_{\rm \mp}$ correspond to the strengths of the sum-frequency components at $2\omega_{\rm p} \mp \omega_{\rm m}$, and are functions of both pump and SH resonance parameters. 

Finally, the total SH light beats at a square-law SH photodetector, and the output photocurrent $i_{\rm PD}$ writes:
\begin{equation}
i_{\rm PD} \approx r_{\rm PD}  \kappa_{\rm ex,b}\hbar 2\omega_{\rm p} |B|^2  \approx r_{\rm PD} \kappa_{\rm ex,b}\hbar 2\omega_{\rm p} (|B_0|^2 +2 \epsilon \operatorname{Re}\{ (B_{\rm -} B_0^* + B_0B_{\rm +}^{*}) e^{i \omega_{\rm m }t}\})
\label{S11}
\end{equation}
where $r_{\rm PD}$ is the responsivity of the photodetector. $\operatorname{Re}\{~\}$ denotes the real part of a complex number. 
In Eq. \eqref{S11}, we have approximated the photon energy of the sum-frequency components at $2\omega_{\rm p} \pm \omega_{\rm m}$ to be the same as the SH component at $2\omega_{\rm p}$, as the modulation frequency is much smaller than the optical carrier  ($\omega_{\rm m} \ll 2\omega_{\rm p}$). The second approximation in Eq. \eqref{S11} is valid due to the weak modulation ($\epsilon \ll 1$).
As such, the photocurrent is mainly comprised of two parts: $|B_0|^2$ term corresponds to the SH power of the pump at $2\omega_{\rm p}$ (DC signal), 
and $2\epsilon\operatorname{Re}\{(B_{-} B_0^* + B_{0} B_+^* ) e^{i \omega_{\rm m }t}\}$ 
term corresponds to the interference of the beating signals between the sum-frequency components at $2\omega_{\rm p} \pm \omega_{\rm m}$ and the SH component at $2\omega_{\rm p}$ (AC signal at modulation frequency $\omega_{\rm m}$). 
By expressing it in the phasor notation, i.e. $B_{-} B_0^* + B_{0} B_+^*  = |B_{-} B_0^* + B_{0} B_+^*| e^{i\varphi}$, where $\varphi$ is its argument, the term
$2\epsilon\operatorname{Re}\{(B_{-} B_0^* + B_{0} B_+^* ) e^{i \omega_{\rm m }t}\}$ can be rewritten as $  2\epsilon|B_{-} B_0^* + B_{0} B_+^*| \cos(\omega_{\rm m }t + \varphi)$. 
Since the VNA measures the response at the modulation frequency $\omega_{\rm m}$ it sent out, the amplitude of the VNA transfer function $|H(\omega_{\rm m})|$ is given by: 
\begin{equation}
\begin{aligned}
|H(\omega_{\rm m})| & \sim 4\epsilon\kappa_{\rm ex,b}\hbar \omega_{\rm p}|B_{-} B_0^* + B_{0} B_+^*| 
= 2\epsilon \bigg|\frac{B_{-}}{B_0} + \big(\frac{B_+}{B_{0}}\big)^{*}\bigg| P_{\rm SH}\\
% & =  \frac{4\epsilon\hbar \omega_{\rm p}g^2\kappa_{\rm ex,a}^2 \kappa_{\rm ex,b} |s_{\rm in}|^4}
% % {\sqrt{(\frac{\kappa_{\rm a}}{2})^2 + \delta_{\rm a}^2}
% % \sqrt{(\frac{\kappa_{\rm b}}{2})^2 + \delta_{\rm b}^2}
% % \sqrt{(\frac{\kappa_{\rm a}}{2})^2 + (\delta_{\rm a}+\omega_{\rm m})^2}
% % \sqrt{(\frac{\kappa_{\rm b}}{2})^2 + (\delta_{\rm b}+\omega_{\rm m})^2}
% {\sqrt{\big((\frac{\kappa_{\rm a}}{2})^2 + \delta_{\rm a}^2\big)^3
% \big((\frac{\kappa_{\rm b}}{2})^2 + \delta_{\rm b}^2\big)
% \big((\frac{\kappa_{\rm a}}{2})^2 + (\delta_{\rm a}+\omega_{\rm m})^2\big)
% \big((\frac{\kappa_{\rm b}}{2})^2 + (\delta_{\rm b}+\omega_{\rm m})^2 \big)} 
% }\\
% & = \frac{2\epsilon \sqrt{\big((\frac{\kappa_{\rm a}}{2})^2 + \delta_{\rm a}^2\big)
% \big((\frac{\kappa_{\rm b}}{2})^2 + \delta_{\rm b}^2\big)}}
% {\sqrt{\big((\frac{\kappa_{\rm a}}{2})^2 + (\delta_{\rm a}+\omega_{\rm m})^2\big)
% \big((\frac{\kappa_{\rm b}}{2})^2 + (\delta_{\rm b}+\omega_{\rm m})^2 \big)}} P_{\rm SH}
& =2\epsilon\bigg| \frac{\frac{\kappa_{\rm a}}{2} + i\delta_{\rm a}}{\frac{\kappa_{\rm a}}{2} + i\delta_{\rm a}+ i\omega_{\rm m}}
\frac{\frac{\kappa_{\rm b}}{2} + i\delta_{\rm b}}{\frac{\kappa_{\rm b}}{2} + i\delta_{\rm b}+ i\omega_{\rm m}} - 
\frac{\frac{\kappa_{\rm a}}{2} - i\delta_{\rm a}}{\frac{\kappa_{\rm a}}{2} - i\delta_{\rm a}+ i\omega_{\rm m}}
\frac{\frac{\kappa_{\rm b}}{2} - i\delta_{\rm b}}{\frac{\kappa_{\rm b}}{2} - i\delta_{\rm b}+ i\omega_{\rm m}}
\bigg|P_{\rm SH} \\
&= 
\frac{4\epsilon\omega_{\rm m} 
\sqrt{\big(\delta_{\rm a}^2 \delta_{\rm b} + \delta_{\rm b}^2\delta_{\rm a} + \delta_{\rm a} (\frac{\kappa_{\rm b}}{2})^2 + \delta_{\rm b} (\frac{\kappa_{\rm a}}{2})^2\big)^2 + \big( \delta_{\rm a}\frac{\kappa_{\rm b}}{2} + \delta_{\rm b}\frac{\kappa_{\rm a}}{2} \big)^2\omega_{\rm m}^2}
P_{\rm SH}}
{\sqrt{(\frac{\kappa_{\rm a}}{2})^2 + (\delta_{\rm a}+\omega_{\rm m})^2}\sqrt{
(\frac{\kappa_{\rm b}}{2})^2 + (\delta_{\rm b}+\omega_{\rm m})^2}\sqrt{
(\frac{\kappa_{\rm a}}{2})^2 + (\delta_{\rm a}-\omega_{\rm m})^2}\sqrt{
(\frac{\kappa_{\rm b}}{2})^2 + (\delta_{\rm b}-\omega_{\rm m})^2
}} 
\label{S12}
\end{aligned}
\end{equation}
Eq. \eqref{S12} explicitly describes the amplitude response measured in the VNA. The amplitude is proportional to the out-coupled SH power, and is weak when the modulation frequency approaches zero ($|H(\omega_{\rm m} \rightarrow 0)| \sim 0$). Although Eq. \eqref{S12} provides a thorough expression, it hinders the intuitive representation of the amplitude response. 
We note that in the experiment, the resonance is accessed by tuning the pump frequency close to the resonance frequency (from blue to red). 
While the resonance always remains at the red side of the pump ($\delta_{\rm a}  <0$), otherwise being thermally unstable. 
As such, the lower sideband $\omega_{\rm p} -\omega_{\rm m}$ is more effectively coupled into the microresonator than the upper sideband $\omega_{\rm p} +\omega_{\rm m}$, which is also clear from Eq. \eqref{S9}. This will also cause the amplitude difference between the sum-frequency components $B_{\mp}$. Given the fact that in most cases the SH resonances are also red-detuned with respect to the SH frequency ($\delta_{\rm b} <0 $), the amplitude of $B_{-}$ is generally much larger than $B_{+}$. 
To give a more clear intuition, while without loss of proper approximation, we neglect the beating signal contributed from the sum-frequency component $B_{+}$. The amplitude response in VNA can be approximated as:
\begin{equation}
|H_{\rm approx}(\omega_{\rm m})|  \sim 2\epsilon\bigg| \frac{\frac{\kappa_{\rm a}}{2} + i\delta_{\rm a}}{\frac{\kappa_{\rm a}}{2} + i\delta_{\rm a}+ i\omega_{\rm m}}
\frac{\frac{\kappa_{\rm b}}{2} + i\delta_{\rm b}}{\frac{\kappa_{\rm b}}{2} + i\delta_{\rm b}+ i\omega_{\rm m}} 
\bigg|P_{\rm SH} 
= 
2\epsilon \frac{\sqrt{(\frac{\kappa_{\rm a}}{2})^2 + \delta_{\rm a}^2}\sqrt{
(\frac{\kappa_{\rm b}}{2})^2 + \delta_{\rm b}^2}}
{\sqrt{(\frac{\kappa_{\rm a}}{2})^2 + (\delta_{\rm a}+\omega_{\rm m})^2}\sqrt{
(\frac{\kappa_{\rm b}}{2})^2 + (\delta_{\rm b}+\omega_{\rm m})^2}} P_{\rm SH}
\label{S13}
\end{equation}
It can be seen that, when the total loss rates of the pump and SH modes ($\kappa_{\rm a}$ and $\kappa_{\rm b}$) are relatively small compared to the effective detuning difference $|\delta_{\rm b} - \delta_{\rm a}|$, the VNA response is approximately maximized at the modulation frequencies $\omega_{\rm m} = -\delta_{\rm a}$ or  $\omega_{\rm m} = -\delta_{\rm b}$. 
Such condition generally holds when the pump laser is further detuned after SH is generated. In the experiment, we are able to identify the two peaks in the VNA responses for SH1-4 within the measurement range of the photodetector. While the frequency locations of the two peaks approximately denote the effective detunings of the pump and SH mode resonances, respectively. 

% FIGURE WITH RESONANCE THERMAL SHIFTS
\begin{figure}[htp]
  \renewcommand{\figurename}{Supplementary Figure}
  \centering{
  \includegraphics[width = 0.9\linewidth]{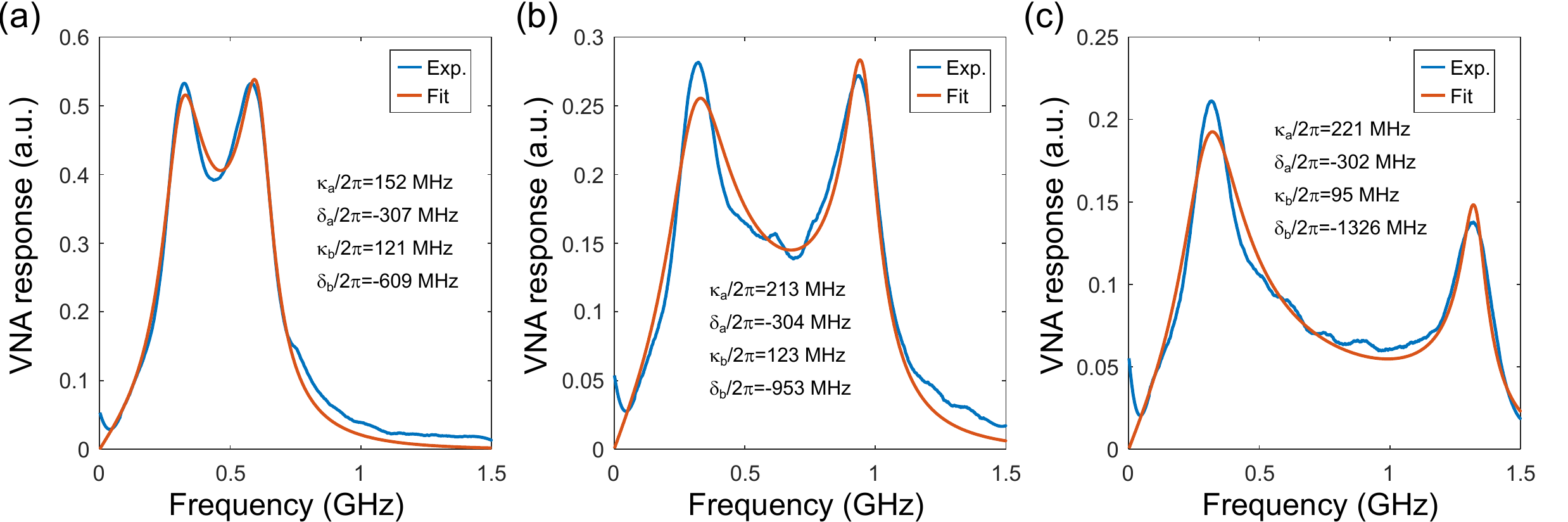}
  } 
  \caption{\noindent\textbf{Fitting of the measured VNA responses. a-c} Experimentally measured (blue) and fitted (orange) VNA responses at pump wavelengths of $1548.62~{\rm nm}$, $1548.64~{\rm nm}$ and $1548.66~{\rm nm}$, respectively. The fitting function is based on Eq. \eqref{S12} and taking into account of the response function of photodetector. The fitted parameters are indicated for each case.   
  } 
 \label{FigureS3}
\end{figure} 

Supplementary Figure \ref{FigureS3} shows three slices of the measured VNA map of Figure 1e in the main text, at pump wavelengths of $1548.62~{\rm nm}$, $1548.64~{\rm nm}$ and $1548.66~{\rm nm}$, respectively.
In order to verify the validity of the derived VNA response, the experimentally measured VNA responses are fitted based on Eq. \eqref{S12}. In the numerical fitting, we also take into account the response function of photodetector having $1~{\rm GHz}$ bandwidth. 
It can be seen that the fitting curves (orange) match well with the measured responses (blue), and their corresponding fitted parameters are indicated for each pump frequency. 
As manifested in Supplementary Figure \ref{FigureS3}, by decreasing the laser frequency, the effective detuning for the pump mode only slightly reduces due to the thermal-induced resonance frequency drag, while the SH cavity resonance walks off quickly. 
Additionally, the fitted effective pump and SH detunings indeed correspond well with the frequencies of the two peaks in a VNA response, as suggested by Eq. \eqref{S13}.

\section*{\textbf{Supplementary Note 3. Probing resonance detunings during AOP events involving multiple SH modes}} 

\begin{figure}[hbt!]
  \renewcommand{\figurename}{Supplementary Figure}
  \centering{
  \includegraphics[width = 0.9\linewidth]{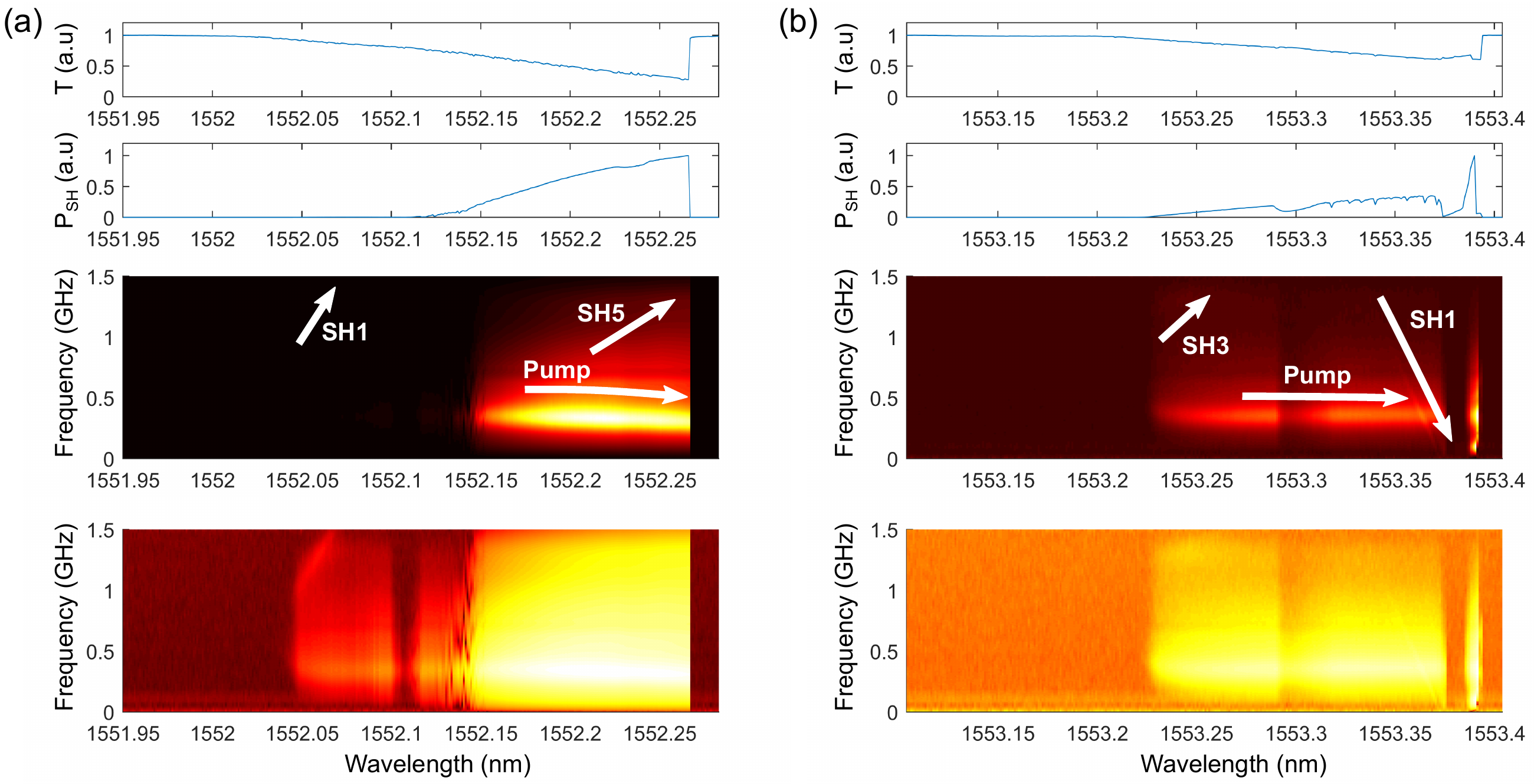}
  } 
  \caption{\noindent \textbf{AOP involving multiple SH modes during the wavelength sweep within a pump resonance.} The measurements were performed in the $146~{\rm GHz}$ microresonator at pump resonances near \textbf{a} $1552.0~{\rm nm}$ and \textbf{b} $1553.2~{\rm nm}$, without external temperature control.
  From top to bottom: normalized pump transmission, generated SH power, measured VNA map in linear scale, and measured VNA map in logarithmic scale. The arrows in the measured VNA maps indicate the movements of pump and SH resonances, where the interacting SH modes are identified using TP imaging. 
  }
 \label{FigureS4}
\end{figure}

In this part, we provide additional VNA measurements in which the AOP involves multiple SH modes within a single pump resonance. 
As shown in Supplementary Figure \ref{FigureS4}a and b, the probing of pump and SH resonance dynamics was carried out in the $146~{\rm GHz}$ microresonator at AOP events near $1552.0~{\rm nm}$ and $1553.2~{\rm nm}$, respectively.
Here we display the normalized pump transmission and generated SH power, as well as the VNA response maps during the wavelength sweeps. 
The VNA maps are plotted in both linear and logarithmic scales for better visibility of  resonance detunings. 
From the measured VNA responses, we could recognize the pump and multiple SH resonance detunings. 
The involved SH modes were identified by TP imaging and indicated in Supplementary Figure \ref{FigureS4}a and b.
For both cases, AOP events were initiated by one SH mode and then subsequently taken over by another SH mode. 
The new $\chi^{(2)}$ gratings arise from the redistribution of space charges to match the most effective doubly resonant condition. 
As evident in Supplementary Figure \ref{FigureS4}b, the AOP resulted from SH1 could even start when the SH resonance is at the blue side of the generated SH ($\delta_{\rm b}>0$). 
This is at the opposite side given the integer number condition of QPM gratings ($\delta_{\rm b} \approx 2\delta_{\rm a} < 0$).
Further, it seems that the $\chi^{(2)}$ grating is erased at the effective zero detuning of the SH resonance ($\delta_{\rm b} \approx 0$), while rewrites after crossing the SH resonance. 
Such phenomenon may be linked to the steep phase slope when exactly on resonance, which prohibits and destructs the $\chi^{(2)}$ grating. 
 
\section*{\textbf{Supplementary Note 4. Image processing and simulation of $\mathbf{ {\chi^{(2)}}}$ gratings}}
\begin{figure}[hbt!]
  \renewcommand{\figurename}{Supplementary Figure}
  \centering{
  \includegraphics[width = 0.55\linewidth]{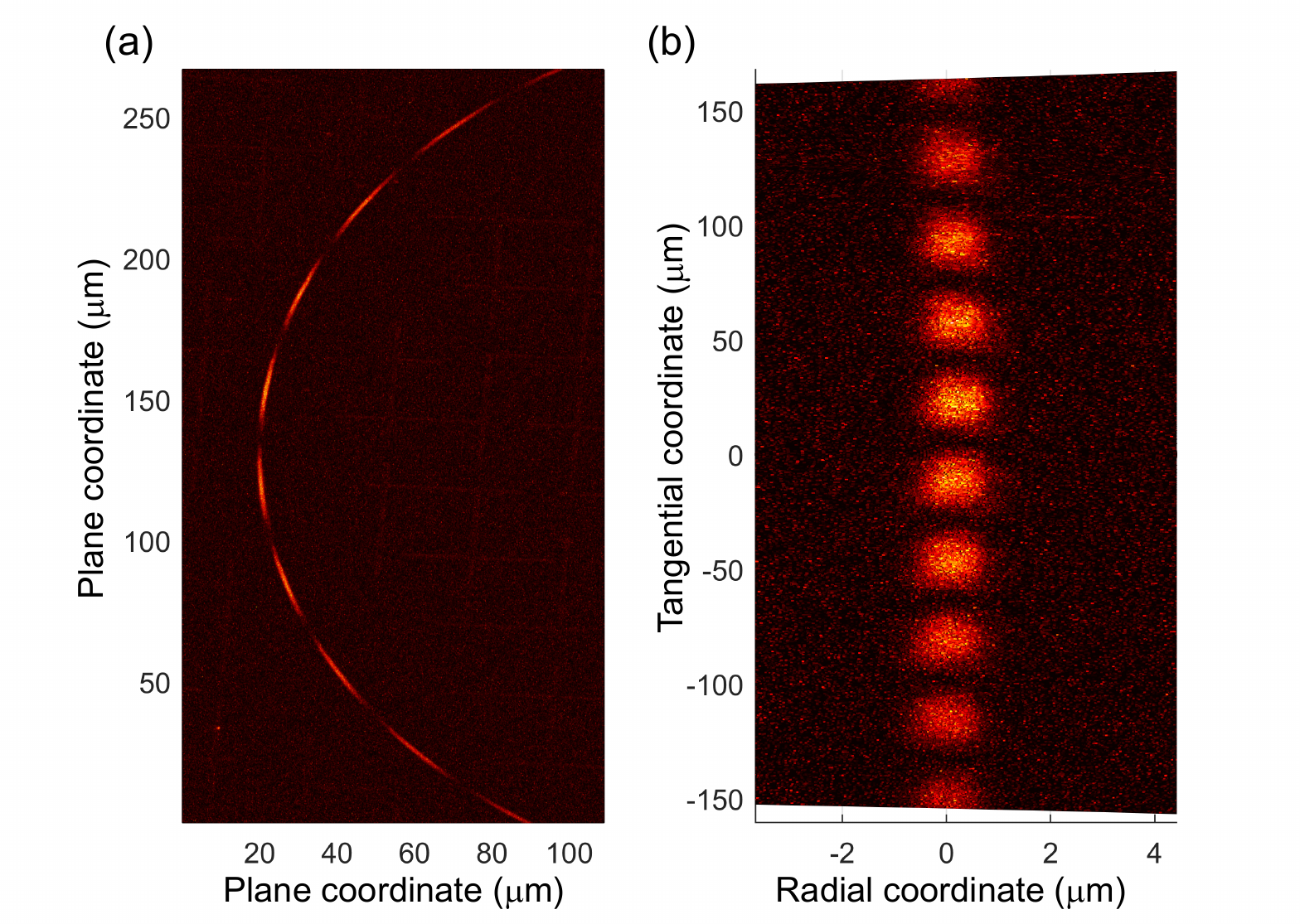}
  } 
  \caption{\noindent\textbf{Two-photon image of the  nonlinear grating in a poled microresonator.} \textbf{a} Two-photon (TP) image of a $146~{\rm GHz}$ Si$_3$N$_4$ microresonator poled at $1559.33 ~{\rm nm}$. The white arrow indicates the incident light polarization. \textbf{b} Replotted image of \textbf{a} in the radial and tangential coordinates of the microresonator.
  }
 \label{FigureS5}
\end{figure}

The measured TP images are replotted in the radial and tangential coordinates of the microresonator to facilitate image processing.  
As an example, Supplementary Figure \ref{FigureS5}a depicts the original TP image when the $146~{\rm GHz}$ microresonator is poled at $1559.33 ~{\rm nm}$, and then replotted in Supplementary Figure \ref{FigureS5}b after coordinate transformation.   
We first identify the center of the microresonator circle $(x_{\rm c}, y_{\rm c})$, and express the points $(x,y)$ of the TP images in the polar coordinate as $(x_{\rm c} + \rho \cos\phi, y_{\rm c} + \rho \sin\phi)$, where $\rho$ and $\phi$ are the radial distance and the azimuthal angle, respectively. Noticeably, $\rho = R$ corresponds to the circumference of the microresonator. Then we map the points $(x_{\rm c} + \rho \cos\phi, y_{\rm c} + \rho \sin\phi)$ of the TP image into points $(\rho, \rho \phi)$ so that the microresonator circumference is transformed into a straight line as shown in Supplementary Figure \ref{FigureS5}b. 
It is worth noting that the TP image records the intensity of SH signal, which is proportional to the square of $\chi^{(2)}$.
Consequently, the observed modulation period in the TP image is half of the inscribed grating period.
The grating shape along the waveguide is simulated as follows. As shown in Supplementary Figure \ref{FigureS1}d, we first simulate the TE-polarized mode field distributions of interacting modes at pump and SH wavelength in the waveguide cross-section plane ($xz$-plane). 
Then we calculate the $\chi^{(2)}$ distribution along the waveguide according to Eq. (3) in the main text. Since the $\chi^{(2)}$ grating is imaged in the $xy$-plane and the recorded TP intensity is proportional to $(\chi^{(2)})^2$, we project the $(\chi^{(2)})^2$ onto $xy$-plane by averaging the values along $z$-direction for the grating shape simulation. The simulated grating shapes are then smoothed using a Gaussian filter so as to take into account the optical resolution of the TP microscope.

%%%%%%%%%%%%%%%%%%%%%%%%%%%%%%%%%%%%%%%%%%%%%
% FIGURE WITH PROCESSING OF TPM IMAGE

\section*{\textbf{Supplementary Note 5. $\chi^{(2)}$ characterization in poled microresonators}}
%%%%%%%%%%%%%%%%%%%%%%%%%%%%%%%%%%%%%%%%%%%%%
% FIGURE WITH GRATING INTENSITY CALIBRATION.
\begin{figure}[hbt!]
  \renewcommand{\figurename}{Supplementary Figure}
  \centering{
  \includegraphics[width = 0.6\linewidth]{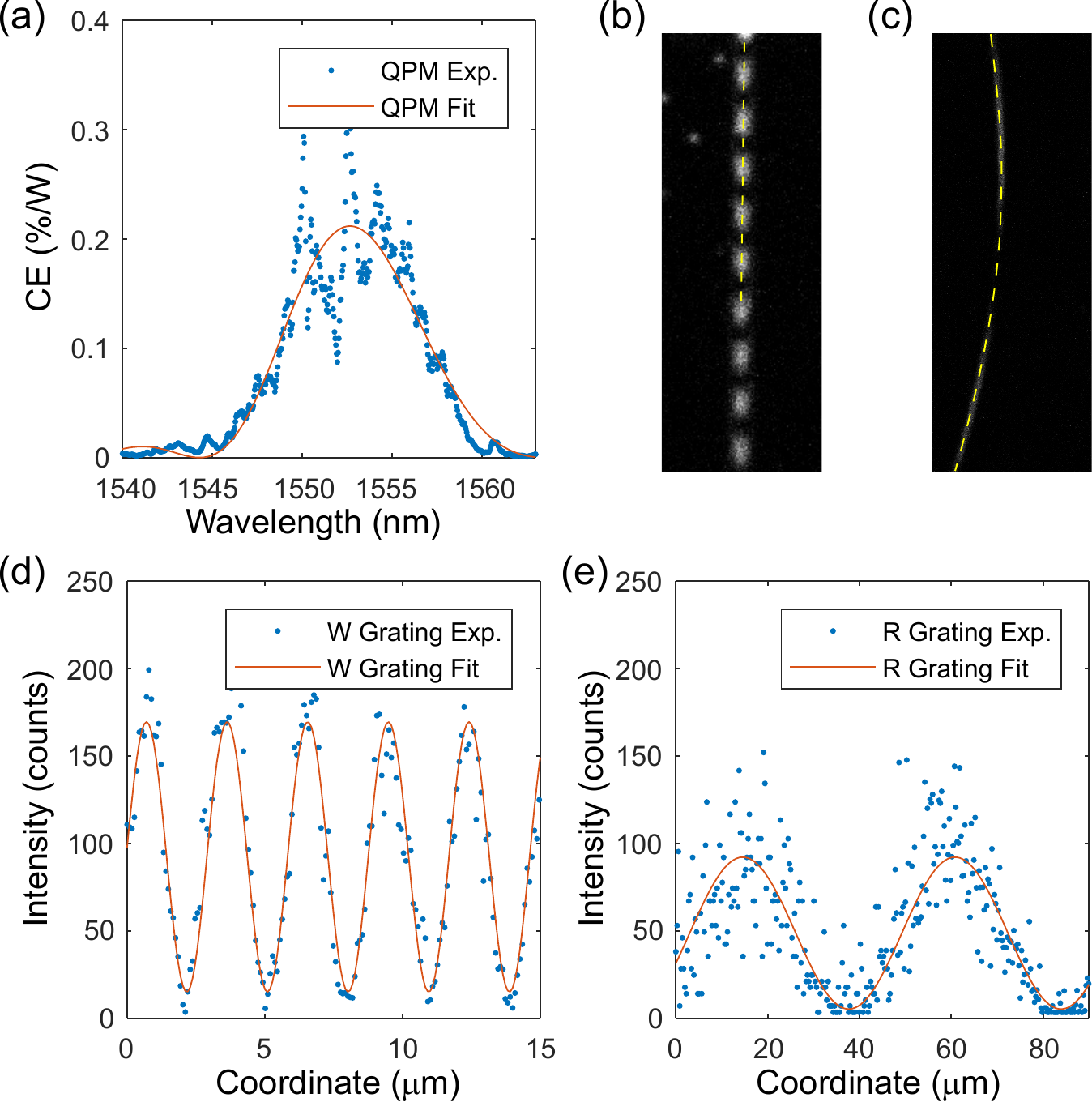}
  } 
  \caption{\noindent\textbf{$\mathbf{\chi^{(2)}}$ characterization in poled straight waveguides and microresonators.}
  \textbf{a} Experimentally measured and fitted conversion efficiency (CE) of a poled Si$_3$N$_4$ waveguide (cross-section $1.8\times0.75~{\rm \mu m^2}$, length $55~{\rm mm}$). $\chi^{(2)}_{\rm W}=0.082~{\rm pm/V}$, $L_{\rm g}=46~{\rm mm}$ and $\Lambda_{\rm W}=5.83~{\rm \mu m}$ are extracted. 
  \textbf{b} TP image of the grating in the straight waveguide with $50\%$ of the available excitation laser power (yellow dashed line denotes the grating extraction direction). \textbf{c} TP image of the grating in the $146~{\rm GHz}$ microresonator poled around $1543.10 ~{\rm nm}$ with $90\%$ of the available excitation laser power (yellow dashed line denotes the grating extraction direction). \textbf{d} Extracted grating intensity (blue) with fitting (orange) along the dashed yellow line in the waveguide (image \textbf{b}). \textbf{e} Extracted grating intensity (blue) with fitting (orange) along the dashed yellow line in the microresonator (image \textbf{c}).
  }
 \label{FigureS6}
\end{figure}

We characterize the strength of $\chi^{(2)}$ being inscribed on the poled microresonators based on TP imaging technique.
% We used TP imaging technique also for the purpose of measuring $\chi^{(2)}$ inside the microresonators after poling. 
Generally, the TP image intensity $I$ of a $\chi^{(2)}$ object is proportional to the square of its $\chi^{(2)}$ as well as the illumination laser power $P$, i.e. $I\sim(\chi^{(2)}P)^2$.
% \begin{equation}
% {I\sim(\chi^{(2)}P)^2}
% \label{S100}
% \end{equation}
Using this relation, we could estimate the inscribed $\chi^{(2)}$ in the microresonator once the measurement system is calibrated via $I$ measurement of a known $\chi^{(2)}$ and $P$. 
For this purpose we used a poled straight waveguide in which the nonlinear grating $\chi^{(2)}$ was determined experimentally based on the SH spectrum. To do so, we first poled a $55~{\rm mm}$ long Si$_3$N$_4$ waveguide (cross-section $1.8\times0.75~{\rm \mu m^2}$ folded into $11$ meanders on a $5\times5~{\rm mm^2}$ chip) employing a nanosecond pulse source\cite{nitiss2019formation}. 
After poling we recorded the conversion efficiency (CE) dependence on the pump wavelength and fitted it with the following function\cite{wang2018ultrahigh, fejer1992quasi}:
\begin{equation}
{\rm CE_{W}}= \frac{(\omega L_{\rm g} \chi^{(2)}_{\rm W})^2}{2 \epsilon_0 c^3 n^2_{\rm eff,a} n_{\rm eff,b}}
\frac{S_{\rm b}}{S^2_{\rm a}}
\big[\frac{{\rm sin}\big(\frac{\delta k L_{\rm g}}{2}\big)}{\big(\frac{\delta k L_{\rm g}}{2}\big)}\big]^2
\label{S110}
\end{equation}
where $\omega$ is the optical frequency and $L_{\rm g}$ is the grating length. 
%$n_{\rm eff,a}$ and $n_{\rm eff,b}$ are the effective refractive indices at pump and SH wavelength, respectively, $S_{\rm a}$ and $S_{\rm b}$ are the effective mode areas at pump and SH wavelength, respectively, 
$c$ is the speed of light in vacuum, $\epsilon_0$ is the vacuum permittivity, and $\delta k$ is the net wavevector mismatch after QPM compensation, i.e. ${\delta k (\omega)}= {k_{\rm b}(\omega) - 2k_{\rm a}(\omega)  - \frac{2\pi}{\Lambda}}$. $k_{\rm a (b)}$, $n_{\rm eff, a(b)}$, and $S_{\rm a (b)}$ are the propagation constants, effective refractive indices, and the effective mode areas at pump and SH wavelength, respectively. The experimentally retrieved ${\rm CE_{W}}$ curve is shown in Supplementary Fig. \ref{FigureS6}a together with its fitting based on Eq. \eqref{S110}.
From fit we obtain the grating parameters inside the waveguide to be $\chi_{\rm W}^{(2)}=0.082~{\rm pm/V}$, $L_{\rm g}=46~{\rm mm}$ and $\Lambda_{\rm W}=5.83~{\rm \mu m}$. 
In TP imaging the nonlinear grating in the straight waveguide was captured at $P_{\rm W} \sim 50\%$ of available excitation laser power (Supplementary Figure \ref{FigureS6}b). 
Afterwards, the laser power was increased to $P_{\rm R}\sim 90\%$ of available excitation power to capture the grating inside the microresonator being poled at $1543.10~{\rm nm}$ (Supplementary Figure \ref{FigureS6}c). The excitation laser power was adjusted in order to achieve good visibility of the grating in each particular measurement,  while all the rest of instrument parameters (dwell time, pixel size, etc.) remained unchanged.
The intensities of gratings in straight waveguide and microresonator are then extracted and fitted as shown in Supplementary Figure \ref{FigureS6}d and e, respectively, based on the following fitting function:
\begin{equation}
I(x)=I_0 \sin^2(\frac{2\pi x}{\Lambda}+x_0)+ I_1
\label{S120}
\end{equation}
where $x$ is the position along the nonlinear grating. $I_0$, $I_1$, and $x_0$ are the fitting parameters. Hence, the $\chi_{\rm R}^{(2)}$ inside the microresonator can be simply extracted when compared to the straight waveguide:
\begin{equation}
\chi_{\rm R}^{(2)}=\chi_{\rm W}^{(2)}\sqrt{\frac{I_{0,{\rm R}}}{I_{0,{\rm W}} }}\frac{P_{\rm W}}{P_{\rm R}}
\label{S130}
\end{equation}
where $I_{0,{\rm R}}$ and $I_{0,{\rm W}}$ are the fitted grating intensities inside the microresonator and straight waveguide, respectively. 
For this particular case at poling wavelength of $1543.10~{\rm nm}$, we obtain the $\chi^{(2)}$ inside the microresonator to be $\chi_{\rm R}^{(2)}=0.031~{\rm pm/V}$, while the $\chi^{(2)}$ at the other wavelengths are determined in the same manner.   

\bibliographystyle{naturemag}
\bibliography{ref}